\documentclass[Times,twocolumn,tighten,deluxetable]{aastex7}

\usepackage{graphicx}
\usepackage{rotating}
\usepackage{lineno}
\usepackage{multirow}
\usepackage{enumerate}

\submitjournal{ApJ}

\shorttitle{Probing Accretion Flow Dynamics of GS 1354-64}
\shortauthors{Debnath et al.}

\begin{document}
\title{Probing Accretion Disk Dynamics of GS~1354$-$64: Insights from 2025$-$2026 Complex Outburst}

\correspondingauthor{Dipak Debnath}
\email{dipakcsp@gmail.com}

\author[0000-0003-1856-5504]{Dipak Debnath}
\affiliation{Institute of Astronomy, National Tsing Hua University, Hsinchu 300044, Taiwan}
\affiliation{Indian Center for Space Physics, 466 Barakhola, Netai Nagar, Kolkata 700099, India}
\affiliation{Center for Theory-Computation-Data Science Research (CTCD), National Tsing Hua University, Hsinchu 300044, Taiwan}
\email{dipakcsp@gmail.com}

\author[0009-0005-2802-0809]{Rahul Bhowmick}
\affiliation{Institute of Astronomy, National Tsing Hua University, Hsinchu 300044, Taiwan}
\email{rahulbhowmickin@gmail.com}

\author[0000-0002-5617-3117]{Hsiang-Kuang Chang}
\affiliation{Institute of Astronomy, National Tsing Hua University, Hsinchu 300044, Taiwan}
\affiliation{Department of Physics, National Tsing Hua University, Hsinchu 300044, Taiwan}
\affiliation{Center for Theory-Computation-Data Science Research (CTCD), National Tsing Hua University, Hsinchu 300044, Taiwan}
\email{hkchang@mx.nthu.edu.tw}

\author[]{Anuj Nandi}
\affiliation{Space Astronomy Group, ISITE Campus, U. R. Rao Satellite Centre, Outer Ring Road, Marathahalli, Bangalore 560037, India}
\email{anuj@ursc.gov.in}



\begin{abstract}

We present a comprehensive spectro-temporal study of the 2025--2026 outburst of the Galactic black hole X-ray binary GS 1354$-$64 using observations from \textit{MAXI}/GSC, \textit{Insight}-HXMT, and \textit{NuSTAR}. The source underwent a $\sim126$ day outburst characterized by an `unusual' sequence of spectral states driven by three short X-ray flares ($\sim2$, 5, and 7~days), likely triggered by episodic enhancements in the Keplerian flow. These flares produced rapid spectral softening and delayed the outburst decay. The \textit{MAXI}/GSC light curves show soft X-ray dominance during the intermediate $\sim90$ days of the outburst. Corresponding minima in the hardness ratio and forward-to-backward loops in the hardness--intensity diagram indicate repeated spectral softening and episodic changes in the accretion flow. Low-frequency type-C QPOs exhibit a three-phase evolution, with frequencies increasing from $0.07$ to $5.76$~Hz, decreasing from $6.14$ to $3.26$~Hz, and rising again to $6.25$~Hz with a sporadic detection of type-B QPOs in the intermediate state. The propagating oscillatory shock model successfully reproduces this evolution through inward shock propagation during the rising phase, followed by rapid outward and gradual inward motion during plateau-1, tracing changes in the Comptonizing corona. Broadband ($2-70$~keV) spectral analysis reveals coupled evolution of the thermal (\texttt{diskbb}: $T_{\rm in}=0.25$--$1.04$~keV) and Comptonized (\texttt{thcomp}: $\Gamma_{\rm th}=1.69$--$2.66$, $f_{\rm cov}\sim0.02$--$0.97$) components. Relativistic reflection is well described by the \texttt{relline} model, while an additional \texttt{powerlaw} above $\sim50$~keV in the initial hard state suggests a high-energy component, possibly associated with synchrotron cooling. Strong correlations between QPO frequency, photon index, and hardness ratio constrain the evolution of the accretion flow geometry and Comptonizing corona.

\end{abstract}


\section{Introduction}

GS~1354$-$64 (= BW~Cir) is a well-known Galactic transient black hole X-ray binary (BH-XRB), discovered during an outburst in 1987 
by the \textit{Ginga} satellite \citep{Swinbanks87,Makino87}. The source shares its sky position with the previously reported BH 
candidates Cen~X$-$2 \citep{Francey71} and MX~1353$-$64 \citep{Markert79}, which were later identified as the same source \citep{Kitamoto90}.
Later, \citet{Casares09} dynamically confirmed GS~1354$-$64 as a stellar-mass BH X-ray binary with a BH mass of 
$M_{\rm BH}=7.9\pm0.5~M_\odot$, an orbital period of $P_{\rm orb}=2.54$~days, a distance of $D \geq 25$~kpc, and an inclination 
angle of $i\sim75^\circ$. Alternatively, the parallax measurement from \textit{GAIA} DR2 yielded a source distance of $0.7^{+3.4}_{-6.3}$~kpc \citep{Gandhi19}, whereas the \textit{GAIA} EDR3 measurement suggests a distance range of $5$--$20$~kpc at $76\%$ confidence \citep{P21}. The companion star was identified as a G$0$–$5$~III giant with a mass of $1.1\pm0.1~M_\odot$ \citep{Casares04}. 
Using X-ray spectral analysis, \citet{El-Batal16} reported that the BH is rapidly spinning, with a dimensionless spin parameter of $a_k=0.98$.

Recently, GS~1354$-$64 showed a new outburst on 2025 December 25 (MJD~61034), as first reported by \textit{MAXI}/GSC \citep{Negoro25}. 
The outburst has since been extensively monitored across the radio, optical, and X-ray bands. Using \textit{IXPE}, \citet{Ravi26} 
detected significant X-ray polarization with a polarization degree (PD) of $(4\pm0.2)$\% and a polarization angle (PA) of 
$1^\circ\pm2^\circ$ in the $2$--$8$~keV band. A broad Fe~K$\alpha$ emission line was detected with the high-resolution Resolve 
instrument onboard \textit{XRISM} \citep{Liu26}. Bright radio flares with increasing flux during the early rising phase of the 
outburst were also reported \citep{Salvaggio26a,Salvaggio26b,Salvaggio26c}. 
\citet[][hereafter Paper-I]{D26} investigated the early phase of the 2025--2026 outburst of GS~1354$-$64 using archival \textit{MAXI}, \textit{Insight}-HXMT, and \textit{NuSTAR} observations. They traced the evolution of low-frequency Quasi-periodic Oscillations 
(LFQPOs) during the rising phase and detected prominent relativistic reflection features in the broadband spectra. Spectral fitting with the \texttt{relxill} model yielded a near-maximal black hole spin ($a_k\sim0.998$) and an inclination of $i\sim63^\circ$--$70^\circ$, consistent with previous estimates.

Stellar-mass BH-XRBs (e.g., GS~1354$-$64) emit radiation across the electromagnetic spectrum, from radio to high-energy $\gamma$-rays. These emissions originate from different regions of the accretion flow and arise from distinct physical processes associated with mass transfer from the companion star. Based on their long-term X-ray behavior, BH-XRBs are classified as persistent (e.g., Cyg~X$-$1), transient (e.g., GS~1354$-$64), or semi-persistent systems. Persistent sources remain bright, with typical X-ray luminosities of $L_{\rm X}\gtrsim10^{36}~{\rm erg~s^{-1}}$ \citep{Tanaka96}, whereas transients spend most of their lifetime in quiescence ($L_{\rm X}\lesssim10^{33}~{\rm erg~s^{-1}}$; \citealt{Hannikainen05}) and undergo episodic outbursts lasting weeks to months. Semi-persistent systems exhibit transient-like outbursts but remain in a high-luminosity state for years; for example, \object{Swift J1753.5$-$0127} stayed active for nearly five years after its 2005 outburst \citep{Soleri13}.

The mechanism that triggers outbursts in BH-XRBs remains uncertain, although enhanced viscosity in the outer accretion disk is widely considered the primary driver \citep{Ebisawa96}. Extending the limit-cycle model for dwarf novae \citep{Cannizzo93}, \cite{C19} introduced the pile-up radius ($X_p$), where matter accumulates during quiescence. An outburst occurs when the viscosity at $X_p$ exceeds a critical value. Likewise, the disk instability model (DIM; \citealt{Lasota01}) attributes outbursts to the accumulation of cold material in the outer disk until hydrogen ionization triggers a thermal-viscous instability. The pile-up radius paradigm has since successfully explained the onset of outbursts, the transition from quiescence to outburst, and the origin of complete and failed outbursts in several transient recurrent BH-XRBs \citep{C19,Bhowmick21,Chatterjee22}.

Transient BH-XRB outbursts exhibit rapid spectral and timing evolution, typically progressing through the low-hard state (LHS), hard-intermediate state (HIMS), soft-intermediate state (SIMS), and high-soft state (HSS) before returning to the LHS, forming the characteristic hysteresis loop in the hardness--intensity diagram (HID) \citep{Belloni05,RM06,N12,N24,D13,D15,Sreehari19}. Based on this evolution, \cite{D17} classified outbursts into type-I, which traverse all four states, and type-II, in which the HSS and sometimes the intermediate states are absent. Low-frequency QPOs (types A, B, and C; \citealt{Casella05}) provide complementary probes of the accretion flow with type-C QPOs evolve systematically through the LHS and HIMS, whereas the more sporadic type-A and type-B QPOs are typically detected during the SIMS \citep{Belloni05,D08,D10,D13,D25,N12,Radhika16,Choudhury25}. The HID and Accretion Rate Ratio--Intensity Diagram (ARRID) are widely used to trace spectral state transitions and accretion flow evolution \citep[][and references therein]{Fender04,Belloni05,N12,N18,Jana16}. These state transitions are closely linked to jet activity, with compact jets in the LHS/HIMS, episodic jets in the SIMS, and quenched jet emission in the HSS \citep[e.g.,][]{Vadawale01,N01,Fender04,Fender09,Radhika16,Jana17,Chatterjee19,D20,D21,Choudhury25}. An exception is EXO~1846$-$031, where strong jet signatures were detected in the HSS immediately after the SIMS-to-HSS transition, suggesting a non-standard jet origin \citep{Bhowmick25}.

The X-ray spectra of BH-XRBs consist of thermal and non-thermal continua together with reflection features, most notably the Fe K$\alpha$ emission line ($\sim$6.4--6.7 keV) and the Compton reflection hump. The thermal component originates from a geometrically thin, optically thick Keplerian accretion disk \citep{SS73} and is commonly described by a multicolor disk blackbody \citep{Mitsuda84,Makishima86a}. The non-thermal component is produced primarily by inverse-Compton scattering of soft disk photons by `hot' electrons in a Comptonizing region, with synchrotron emission contributing in some cases \citep{ST80,ST85,Wardzinski00}. Several geometries have been proposed for this Comptonizing region, including magnetic and disk coronae \citep{Galeev79,Haardt93,Zdziarski03,Done07} and the two-component advective flow (TCAF) model \citep{CT95}, in which the Compton cloud forms naturally in the post-shock region of the sub-Keplerian flow. In TCAF, its properties are governed by the Keplerian and sub-Keplerian accretion rates, along with the shock location ($X_s$) and strength (i.e., compression  $R$). Reflection features arise from the interaction of hard X-rays with the relatively cold accretion disk through Compton scattering, photoelectric absorption, and fluorescence \citep{Lightman88,Fabian89,Brenneman06,Garcia10,Bambi21}. In particular, the relativistically broadened Fe K$\alpha$ line provides a powerful probe of the inner accretion flow and enables robust measurements of black hole spin \citep{Makishima86b,Fabian89,Garcia10,Reynolds14}.

The outburst properties of transient BH-XRBs including duration, peak luminosity, fluence, flare activity, and morphology vary widely both among sources and between successive outbursts of the same source \citep[e.g.,][]{Chen97,Zhang19}. Some outbursts exhibit particularly complex spectro-temporal evolution, posing challenges for their physical interpretation \citep{Baby20,Nath23}. As reported in Paper-I, the 2025--2026 outburst of GS~1354$-$64 is `unusually' complex owing to multiple short-duration X-ray flares. Motivated with this, we present a comprehensive spectro-temporal analysis of archival \textit{MAXI}, \textit{NuSTAR}, and \textit{Insight}-HXMT observations to investigate the accretion flow evolution and the physical processes driving this `complex' outburst, where an unconventional sequence of spectral state transitions are found. The study of the monotonic evolution of the QPO frequency using the propagating oscillatory shock (POS) model over an extended epoch provides a physical picture of the propagation of the shock wave responsible for the generation of type-C and type-B QPOs during QPO observed phases of the outburst, and offers insights into their possible connection with the accretion flow, corona and jets/outflows. Furthermore, an investigation of the correlations between the temporal parameters (QPO frequency and HR) and the spectral parameter (photon index) provides a deeper understanding of the evolution of the accretion-flow dynamics and the Comptonizing corona.

The paper is organized as follows: Section~2 describes the observations, data reduction, and analysis procedures. The timing and spectral
results are presented in Section~3, followed by a discussion of their physical implications and conclusions in Section~4.

\section{Observation and Data Analysis}

We analyze on-demand {\it MAXI}/GSC data together with publicly available {\it NuSTAR}/FPMA and {\it Insight}-HXMT/LE, ME, and HE 
observations to investigate the spectro-temporal evolution of GS~1354$-$64 during its 2025--2026 outburst. 
The {\it NuSTAR}\footnote{\url{https://heasarc.gsfc.nasa.gov/docs/nustar/analysis}} and 
{\it Insight}-HXMT\footnote{\url{http://hxmten.ihep.ac.cn/analysis.jhtml}} data were reduced using the standard analysis procedures 
and the latest HEASoft package (v6.36). {\it NuSTAR} data reduction was performed through the SciServer 
platform\footnote{\url{https://www.sciserver.org}}, while {\it Insight}-HXMT data were processed locally after downloading 
the archival observations.

\subsection{MAXI}

We downloaded the daily-averaged on-demand {\it MAXI}/GSC fluxes from the MAXI RIKEN on-demand website\footnote{\url{https://maxi.riken.jp/mxondem}} 
for the period 2025 December 25 (MJD 61034) to 2026 May 6 (MJD 61166). The data cover the soft (2--6 keV) and hard (6--20 keV) X-ray bands, 
and their flux ratio is defined as the hardness ratio (HR1). The soft, hard, and total (2--20 keV) fluxes were also converted to Crab units 
using the appropriate conversion factors obtained from HEASARC's WebPIMMS\footnote{\url{https://heasarc.gsfc.nasa.gov/cgi-bin/Tools/w3pimms/w3pimms.pl}}.

\subsection{NuSTAR}

We use publicly available eleven {\it NuSTAR} \citep{Harrison13} TOO observations obtained between 2026 January 13 (MJD 61053.34) and April 29 
(MJD 61159.13). The data were reduced using the latest {\fontfamily{qcr}\selectfont NuSTARDAS} version (v2.1.4) and calibration file (CALDB20260527). 
Cleaned event files were generated using the {\fontfamily{qcr}\selectfont nupipeline} task.

Source light curves (with 10 s and 0.01 s time bins) and spectral products (PHA, ARF, and RMF) were extracted using a circular region of 
radius $50''$ centered on the source position. Background products were obtained from an annular region with inner and outer radii of 
$180''$ and $240''$, respectively. The {\fontfamily{qcr}\selectfont nuproducts} task was used for the extraction of both light curve 
and spectral files. The spectra were regrouped using {\fontfamily{qcr}\selectfont GRPPHA} with a minimum of 100 counts per bin to improve 
the signal-to-noise ratio (SNR).

\subsection{Insight-HXMT}

We use publicly available archival 115 observations from the {\it Insight}-HXMT/LE, ME, and HE instruments \citep{Zhang14,Zhang20}, obtained 
between 2026 January 18 (MJD 61058.18) and April 1 (MJD 61131.68). These observations are used to study the outburst profile in the soft 
($2$--$6$~keV), hard ($6$--$50$~keV), and total ($2$--$50$~keV) energy bands. The hard and soft band count rates are further used to 
investigate the evolution of hardness ratio (HR2), defined as their ratio.

We used the latest calibration files (v2.08) and the {\tt hpipeline} pipeline within the HEASoft package to extract temporal and spectral 
products from the three instruments. Light curves with $10$~s and $0.01$~s time bins were generated in the LE ($2$--$6$, $6$--$10$, and 
$2$--$10$~keV), ME ($10$--$20$, and $10$--$30$~keV), and HE ($30$--$50$~keV) energy bands, while spectra were extracted over the default energy ranges of 
the three instruments. To improve the SNR, the LE and ME spectra were rebinned to a minimum of $50$ counts per bin and the HE spectra to 
at least $30$ counts per bin using the {\fontfamily{qcr}\selectfont GRPPHA} task. 
Due to the low SNR ($< 3~\sigma_{\rm SNR}$) and background dominance above $\sim7$~keV, observations obtained after 2026 March 07 (MJD~61107) 
were excluded from the spectral analysis. In addition, $5$ observations acquired before MJD~61107 were also not considered for spectral study because the corresponding LE spectra were unavailable. Consequently, a total of $75$ observations were included in the spectral analysis. These observations without LE data were, however, used exclusively for the timing analysis (see Table~\ref{table_master}).

\subsection{Data Analysis}

The daily average fluxes {\it MAXI}/GSC and rates of {\it Insight-HXMT}/LE, ME, HE instruments are used to study nature of the 
outburst in hard, soft and total energy bands. The ratio of hard to soft energy band fluxes/rates of these two observatories are 
also used to study the evolution of X-ray color-color or hardness-ratios (HRs) and hardness-intensity diagrams (HIDs). 

Fast Fourier transformed power density spectra (PDS) using $0.01$~s time-binned light curves from {\it Insight-HXMT} and 
{\it NuSTAR} observations are used to find one of the important temporal features of BH-XRBs, namely QPOs. The adopted time resolution corresponds to a Nyquist frequency of $50$~Hz, which is adequate to search for LFQPOs, as no HFQPOs are detected in our observations. 
The {\fontfamily{qcr}\selectfont powspec} task of HEASoft subpackage {\fontfamily{qcr}\selectfont XRONOS} with 
{\fontfamily{qcr}\selectfont norm=2} is used to generate PDS in the \textit{Insight-HXMT} combined LE plus ME band of 
$2-20$~keV (or $10-20$~keV ME band where LE data are unavilable) or \textit{NuSTAR}/FPMA energy band of $3-20$~keV. 
To understand nature of the QPOs, PDS are fitted with combination of \textit{Lorentzian} and \textit{constant} models. 
The QPO centric \textit{Lorentzian} model fit provides centroid frequency, full width at half maximum (FWHM), and power of the QPOs.
From these best-fitted Lorentzian parameters, we derived the quality factor ($Q=\nu/\mathrm{FWHM}$), the fractional rms amplitude 
(\%rms $=100\sqrt{\pi \times \mathrm{power} \times \mathrm{FWHM}/2}$), and the detection significance ($\sigma$), 
defined as the ratio of the best-fit Lorentzian normalization to its negative uncertainty \citep{vanderKlis94,Belloni02,Belloni12,Sreehari19}.

To investigate the evolution of the thermal and non-thermal emission during the 2025--2026 outburst of GS~1354$-$64, we fitted the broadband {\it NuSTAR} and {\it Insight}-HXMT spectra obtained between 2026 January 13 (MJD 61053.60) and April 29 (MJD 61159.30) using three spectral models: M1: {\fontfamily{qcr}\selectfont tbabs(thcomp$\otimes$diskbb)}, M2: {\fontfamily{qcr}\selectfont tbabs(thcomp$\otimes$diskbb+relline)}, and M3: {\fontfamily{qcr}\selectfont tbabs(thcomp$\otimes$diskbb+relline+powerlaw)}. The first ten {\it NuSTAR} observations (MJD 61053.60--61117.50) exhibit a prominent Fe K$\alpha$ emission line near 6.4 keV and relativistic reflection features in the 10--30 keV band. In addition, the first three observations, obtained during the rising phase show excess emission above $\sim50$ keV, requiring the additional \texttt{powerlaw} component. Consequently, the first three {\it NuSTAR} spectra were fitted with M3, the remaining eight reflection-dominated {\it NuSTAR} spectra with M2, and all {\it Insight}-HXMT spectra with M1. We fixed the hydrogen column density at $N_{\rm H}=7\times10^{21}~\rm cm^{-2}$ \citep{El-Batal16}. For the \texttt{relline} component, the inner radius, outer radius, spin, and inclination were fixed at $R_{\rm in}=-1~r_g$, $R_{\rm out}=400~r_g$, $a_k=0.998$, and $i=66^\circ$, respectively, where gravitational radius $r_g=GM/c^2$. The adopted spin and inclination correspond to the estimated and average values, respectively, obtained in Paper-I.

Furthermore, after obtaining the best-fitting with the model (M1, M2, or M3) based on an acceptable reduced chi-square ($\chi_\nu^2 \approx 1$), the bolometric flux as well as the \texttt{diskbb} and \texttt{thcomp} component fluxes were estimated over the extended $1$--$100$~keV energy band using the \texttt{cflux} convolution model within the XSPEC package of HEASoft. Notably, low SNR above $\sim$40~keV in many \textit{Insight}-HXMT observations, prevents us from reliably constraining the electron temperature ($kT_{\rm e}$) in the \texttt{thcomp} component. For these observations, after obtaining the best fit with the M1, M2, or M3 model, we freeze $kT_{\rm e}$ at its lower limit, below which fit becomes insensitive to its value (see, Table~\ref{table_master}).

\begin{figure*}
  \centering
    \includegraphics[angle=0,width=18.8cm,keepaspectratio=true]{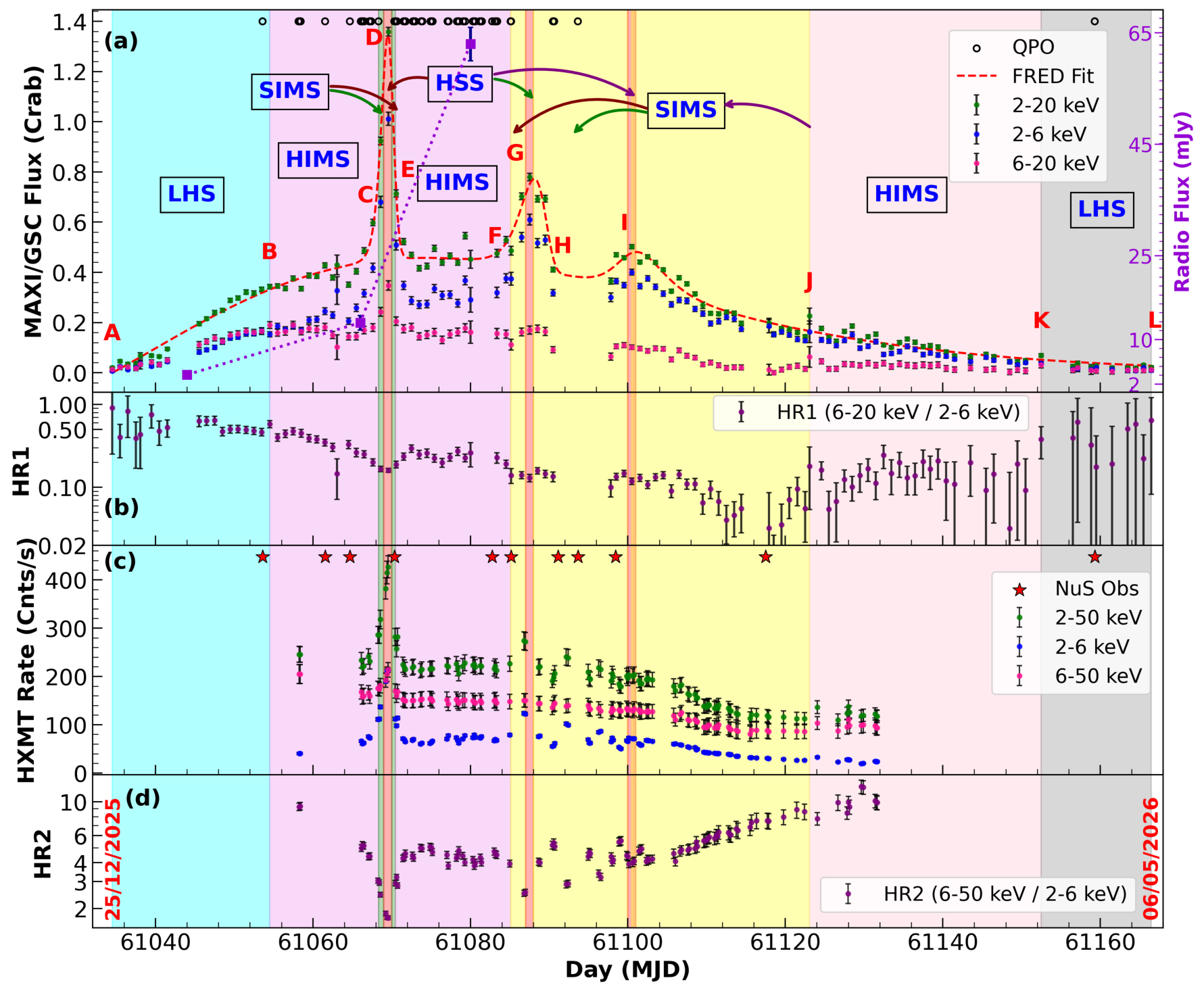}
    \caption{Variation of (a) \textit{MAXI}/GSC soft ($2$--$6$~keV), hard ($6$--$20$~keV), and total ($2$--$20$~keV) fluxes (Crab units), and (b) hardness ratio~1 (HR1 = $6$--$20$/$2$--$6$~keV). Panels (c) and (d) show the \textit{Insight}-HXMT soft ($2$--$6$~keV), hard ($6$--$50$~keV), and total ($2$--$50$~keV) count rates, and hardness ratio~2 (HR2 = $6$--$50$/$2$--$6$~keV). Different colored bands denote the spectral states identified from the spectro-temporal analysis, while points A--L mark the start, end, and transitions between different spectral states. The red dashed curve represents the combined fit of the four-FRED model to the total flux. The top panel also includes the MeerKAT 1.28~GHz radio fluxes (violet points; \citealt{Salvaggio26a,Salvaggio26b,Salvaggio26c}) and observations of QPOs (open circles). In panel (c), the \textit{NuSTAR} observations (NuS Obs) are marked with star symbols.
	}
\label{fig_lc}
\end{figure*}

\section{Results}

We investigate the evolution of temporal and spectral properties of GS~1354$-$64 throughout its 2025--2026 outburst using on-demand {\it MAXI}/GSC data and archival observations from {\it NuSTAR}/FPMA and {\it Insight}-HXMT/LE, ME, and HE. This multi-mission study provides a comprehensive view of the evolution of the source characteristics and the underlying accretion flow dynamics. 
In Paper-I, we presented initial results on the temporal properties of the source during the early phase of the outburst, covering the period up to the end of the second X-ray flare (19 February 2026; MJD 61090), along with its spectro-temporal behaviour at the peak of the first X-ray flare. 
The main results of the present study are presented in the following subsections.

\begin{figure*}
  \centering
    \includegraphics[angle=0,width=16.5cm,keepaspectratio=true]{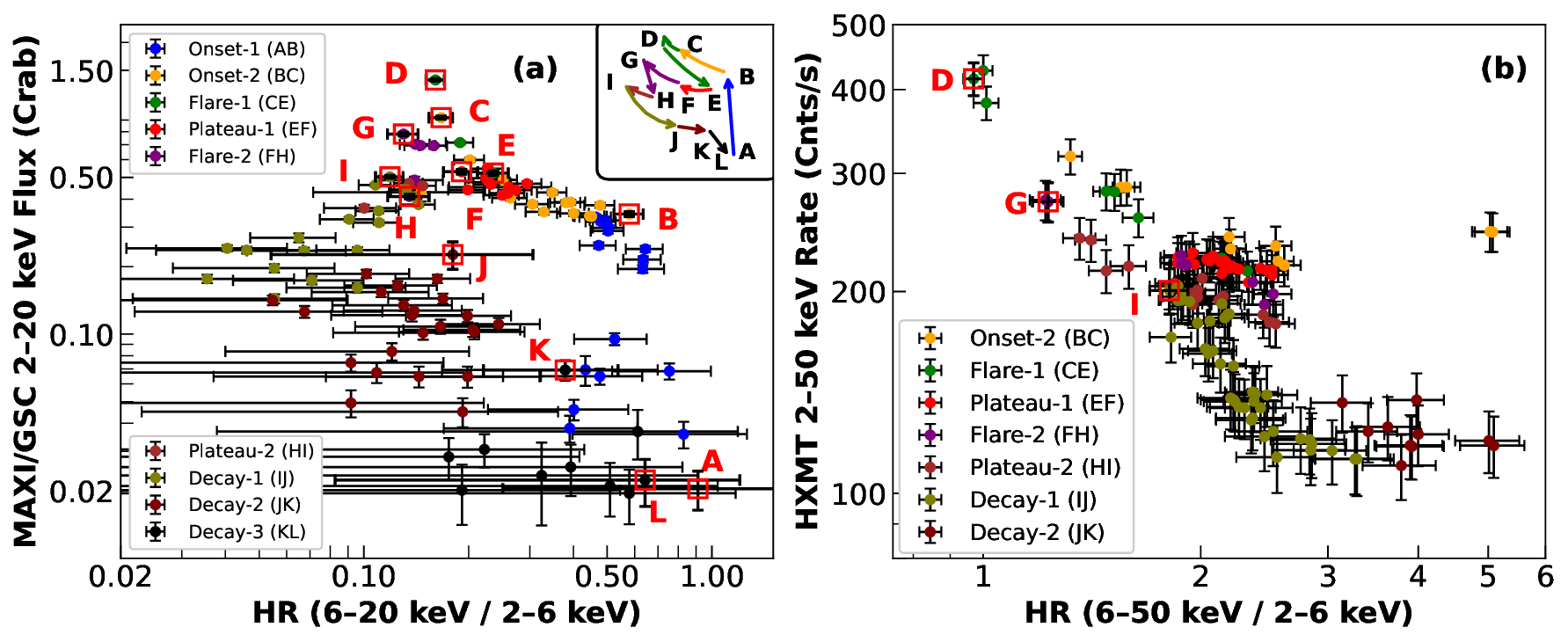}
	\caption{The \textit{MAXI}/GSC hardness--intensity diagram (HID1; left panel) covers the period from 2025 December 25 (MJD~61034) to 2026 May 6 (MJD~61166), while the \textit{Insight}-HXMT HID2 (right panel) spans 2026 January 18 (MJD~61058) to April 1 (MJD~61131). Points A--L mark the different phases of the outburst and the corresponding spectral states. Points D, G, and I indicate the peaks of the three X-ray flares, where the source found to be in a locally soft state (HSS). See text for details.
	}
\label{fig_HID}
\end{figure*}

\subsection{Outburst Profile} 

{\it MAXI} and {\it Insight}-HXMT monitored GS~1354$-$64 throughout its 2025--2026 outburst with an approximately daily cadence. {\it MAXI} covered the entire outburst from 2025 December 25 (MJD~61034) to 2026 May 06 (MJD~61166), while {\it Insight}-HXMT observed the source between 2026 January 18 (MJD~61058.18) and April 01 (MJD~61131.68). We use the soft, hard, and total fluxes/count rates from {\it MAXI}/GSC and {\it Insight}-HXMT/LE, ME, and HE to investigate the evolution of the outburst profile and hardness ratios (HRs; see Figure~\ref{fig_lc}). Here, HR1 is defined as the ratio of the {\it MAXI}/GSC hard ($6$--$20$~keV) to soft ($2$--$6$~keV) fluxes, and HR2 as the ratio of the combined {\it Insight}-HXMT (LE+ME+HE) hard ($6$--$50$~keV) count rate to the LE soft ($2$--$6$~keV) count rate.

To better understand the nature of the outburst, we fitted the outburst profile in the total energy band ($2$--$20$~keV) using a combination of four fast-rise exponential-decay (FRED) profiles \citep{Kocevski03}. 
Although the outburst rise gradually, its later phase of evolution was marked by three short-duration X-ray flares. The peak total band fluxes measured with {\it MAXI}/GSC reached approximately $1.4$, $0.8$, and $0.5$~Crab. In Figure~\ref{fig_lc}(a), points 
A--L mark the onset, end, and major transition phases of the outburst. The intervals C--E (MJD~61068.33--61070.48) and F--H (MJD~61084.51--61091.51) 
correspond to the first and second flares, lasting $\sim2$ and $\sim5$~days, respectively, while points D (MJD~61069.57), G (MJD~61087.51), 
and I (MJD~61100.51) denote the peaks of the three flares.
However, the peak MJDs obtained from the FRED model fits for flares 2 \& 3 differ by $\sim0.6$ and $\sim0.9$~day, respectively, while the peak MJD value of flare-1 remains unchanged.
The {\it MAXI}/GSC light curves show that the soft ($2$--$6$~keV) band dominates both outburst evolution and the hardness ratio, whereas 
the hard ($6$--$20$~keV) flux remains nearly constant except during the first flare, when it increases significantly. {\it Insight}-HXMT 
exhibits a similar evolution in its energy resolved light curves and hardness ratios.

Based on the evolution of the energy-resolved fluxes/rates and hardness ratios from both missions, we identify the canonical spectral states 
(LHS, HIMS, SIMS, and HSS), which are later confirmed by the behavior of QPOs and spectral parameters (see \S3.3 and \S3.4). During the rising phase, the source 
follows the typical sequence ($\mathrm{LHS} \rightarrow \mathrm{HIMS} \rightarrow \mathrm{SIMS} \rightarrow \mathrm{HSS}$) observed in 
classical type-I outbursts of transient BH-XRBs. After the first X-ray flare, the X-ray flux declines rapidly. During the E--F interval 
(MJD~61070.48–61084.51) between flares 1 and 2, except for the initial $\sim1.5$ days, the fluxes and hardness ratios remain nearly 
constant, we designate this interval as plateau-1.

During the X-ray flare peaks, the soft-band flux (or count rate) rises more rapidly than the hard-band emission, producing brief 
($\sim1$ day) local minima in the hardness ratios (see Figure~\ref{fig_lc}(b, d)). We identify these intervals as probable HSS episodes.
The behavior between flares 2 and 3 is poorly constrained by {\it MAXI}/GSC because of a $\sim6$ days data gap. However, {\it Insight-HXMT} observations indicate that during plateau-2 (H--I), the $6$–$50$~keV count rate remains nearly constant, whereas the $2$–$6$~keV rate shows stochastic variations. Thereafter, from I to L, the source follows the canonical decay track of transient BH-XRBs, evolving from softer to harder spectral states ($\mathrm{HSS} \rightarrow \mathrm{SIMS} \rightarrow \mathrm{HIMS} \rightarrow \mathrm{LHS}$). 
Notably, the \textit{Insight}-HXMT soft-band flux shows a positive fluctuation between flares 2 and 3, around MJD~61090, which could indicate an additional flare peak. However, in the absence of simultaneous \textit{MAXI}/GSC observations, we cannot confirm whether this feature represents a distinct flare.

Figure~\ref{fig_lc}(a) also shows the evolution of the $1.28$~GHz MeerKAT radio fluxes reported by \citet{Salvaggio26a,Salvaggio26b,Salvaggio26c}. 
The radio flux increases with the early X-ray rise, indicating enhanced jet/outflow activity. However, the absence of simultaneous radio 
observations during the X-ray flares prevents us from constraining the radio emission behavior throughout these episodes.

\subsection{Hardness-Intensity Diagrams}

The Hardness--Intensity Diagrams (HIDs) provide valuable insight into the accretion flow evolution. Figure~\ref{fig_HID}(a--b) shows the HIDs 
constructed using HR1 (the \textit{MAXI}/GSC $6$--$20$~keV to $2$--$6$~keV flux ratio; HID1) and HR2 (the \textit{Insight}-HXMT $6$--$50$~keV 
to $2$--$6$~keV count-rate ratio; HID2). Owing to the continuous \textit{MAXI}/GSC monitoring throughout the 2025--2026 outburst of GS~1354$-$64, 
the HID clearly correlates the different outburst phases with the branches marked A--L. Despite the increased complexity introduced by repeated 
X-ray flares and the `possibly' replenishment of accreting matter, the overall evolution follows the canonical `q'-shaped HID track, commonly observed in transient BH-XRBs.

The inset of Figure~\ref{fig_HID}(a) illustrates the overall HID evolution track during the outburst. The nearly vertical branches (A--B and K--L), marked by large intensity changes with less variation in hardness, correspond to the rising and declining LHS, respectively. In contrast, the nearly horizontal branches (B--C, E--F, and J--K), characterized by significant hardness changes at nearly constant intensity, represent the rising, plateau, and declining phases of HIMS.
The branches associated with the rising and declining phases of the X-ray flares (C--D--E, F--G--H and H--I--J) exhibit rapid 
forward-to-backward excursions in HR, accompanied by pronounced variations in the soft ($2$--$6$~keV) and total ($2$--$20$~keV) 
\textit{MAXI}/GSC intensities. These rapid transitions make the 2025--2026 outburst of GS~1354$-$64 significantly more complex than 
the evolution of HID, typically observed in transient BH-XRBs. Both \textit{MAXI}/GSC and \textit{Insight}-HXMT HIDs (Figure~\ref{fig_HID}(a,b)) 
show distinct local minima in HR at the peaks of the three X-ray flares (points D, G, and I), indicating short spectral softening episodes 
during which the source approaches the HSS. Excluding these short-lived HSS intervals, the rising and declining flare branches, together 
with the I--J branch, are classified as the SIMS.

The \textit{Insight}-HXMT HID (HID2) does not exhibit a complete canonical `q'-shaped track because observations during the initial and 
late hard states are unavailable. Nevertheless, like HID1, it shows forward-to-backward evolution in HR during the X-ray flares.

\subsection{Evolution of Low Frequency QPOs}

\begin{figure*}
  \centering
    \includegraphics[angle=0,width=17.8cm,keepaspectratio=true]{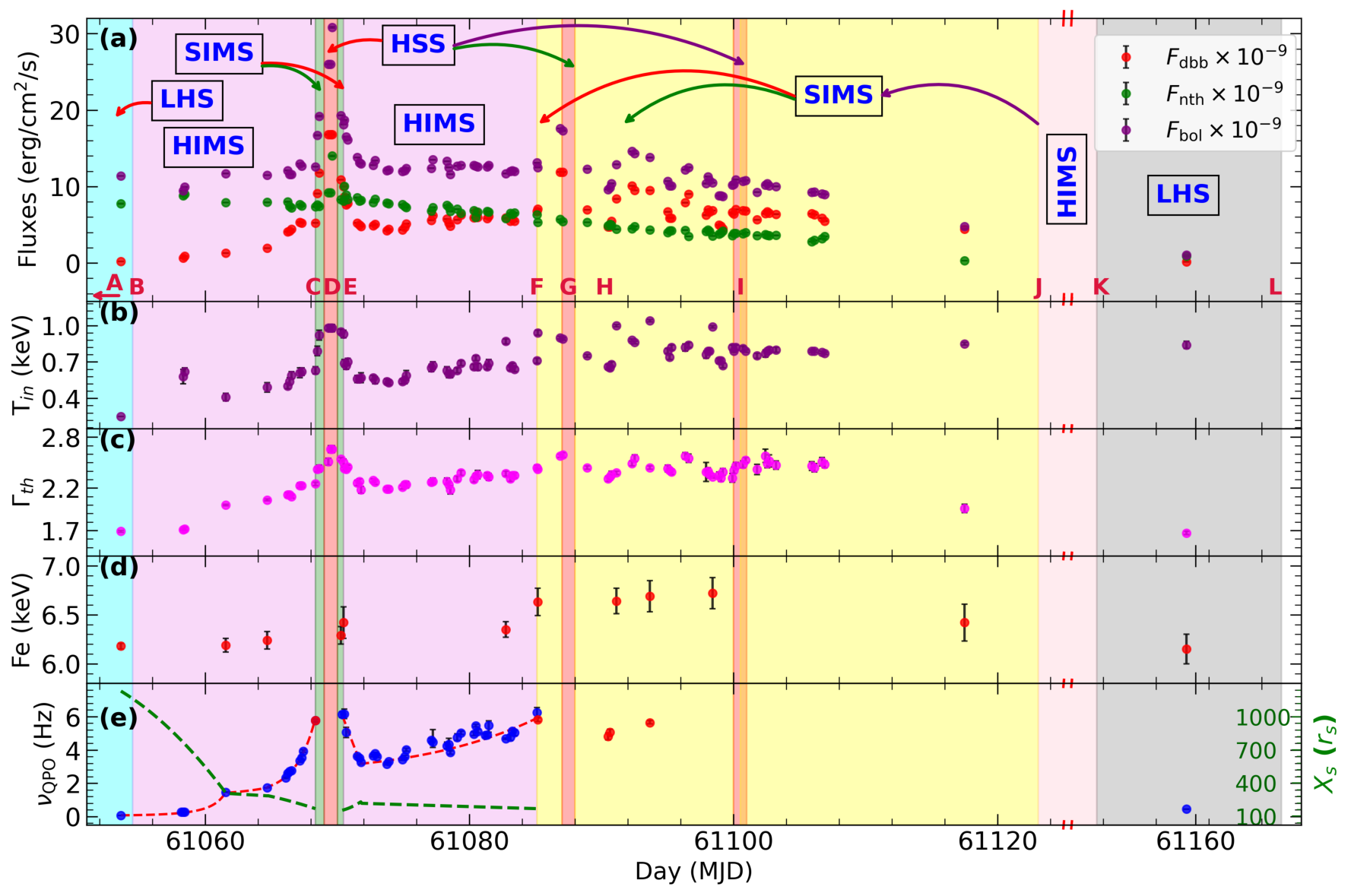}
	\caption{Variation of (a) the bolometric, \texttt{diskbb}, and \texttt{thcomp} fluxes (in units of $10^{-9}~{\rm erg~cm^{-2}~s^{-1}}$) estimated over the $1$--$100$~keV band, (b) the inner disk temperature ($T_{\rm in}$), (c) the \texttt{thcomp} photon index ($\Gamma_{\rm th}$), and (d) Fe line energy, obtained from the best-fit spectral model (M1, M2, or M3) . Panel (e) shows the evolution of the QPO frequency together with the best-fit POS model (red dashed curve). The corresponding shock location inferred from the POS model is shown by the green dashed curve. Blue and red points represent type-C and type-B QPOs, respectively. See text for details.
    }
\label{fig_spec_para}
\end{figure*}

High time-resolution observations from \textit{NuSTAR} and \textit{Insight}-HXMT were used to investigate LFQPOs, a key timing signature of BH-XRBs. Significant QPOs were detected in 52 of the 91 observations. Their evolution during the rising and first plateau phases was presented in Paper-I, where the monotonic frequency evolution before and after the first X-ray flare was successfully explained by the propagating oscillatory shock (POS) model \citep{C08,D10,D13,D25,Iyer15}, constraining the shock properties within the shock oscillation framework \citep[e.g.,][]{MSC96,RCM97,C15}. Here, we present all detected QPOs, their Lorentzian fits, and derived parameters (see Table~\ref{table_master}). Based on their timing properties and broadband noise characteristics \citep{Casella05}, the QPOs are classified into different types. Type-C QPOs dominate throughout the outburst, except for one observation during the rising SIMS and four during the declining SIMS (red points in the bottom panel of Figure~\ref{fig_spec_para}), where type-B QPOs are detected.

During the rising LHS (A--B) and HIMS (B--C), the type-C QPO frequency increases monotonically from $73$~mHz (MJD~61053.60) to $5.76$~Hz (MJD~61068.33), terminating with the detection of a type-B QPO at $5.76$~Hz, which was not reported in Paper-I. No significant QPOs are detected during the first high-fluence X-ray flare (C--D--E). A type-C QPO reappears at point E (MJD~61070.33) in the \textit{NuSTAR} observation (ObsID 91202308002), becoming detectable only during the final $\sim15$~ks of the $\sim40$~ks exposure (see Table~\ref{table_master}). Its frequency then decreases rapidly from $6.16$ to $3.26$~Hz over the next $\sim1.5$ days (MJD~61070.33--61071.78). During the subsequent $\sim14$ day plateau-1 (HIMS), type-C QPOs persist between $3$ and $6$~Hz, showing an overall increasing trend and reaching $6.25$~Hz on MJD~61085.11 (see bottom panel of Figure~\ref{fig_spec_para}). Over the following $\sim8$ days, four type-B QPOs with frequencies of $4.8$--$5.6$~Hz are detected during the declining SIMS. Finally, a type-C QPO at $\sim0.45$~Hz is detected in the late declining LHS on MJD~61159.30. The absence of observations during the declining HIMS and most of the LHS (branches J--L) prevents tracking the expected monotonic decrease in the type-C QPO frequency.

In Paper-I, the POS model successfully reproduced the monotonic evolution of the type-C QPO frequency up to MJD~61071.78. Here, we revise the analysis by including the newly identified 5.76~Hz QPO immediately preceding the first X-ray flare. The updated fit shows that, during the rising phase (A--C) the shock propagated inward from $1230~r_s$ to $166~r_s$ with a velocity of $\sim0.2$--$2~{\rm m~s^{-1}}$, while its compression ratio gradually decreased. After the first flare, the shock receded from $151~r_s$ to $228~r_s$ at an approximately constant velocity of $\sim0.8~{\rm m~s^{-1}}$, producing the observed decline in QPO frequency.
We further extend the POS model analysis to the type-C QPOs detected through 2026 February 14 (MJD~61085.11). Although the QPO frequency during plateau-1 is not strictly monotonic, its overall increase is well reproduced by continued inward shock propagation at a much lower velocity of $\sim0.1~{\rm m~s^{-1}}$. Over this $\sim14$ day interval, the shock moves from $228~r_s$ to $166~r_s$ (see Figure~\ref{fig_spec_para}) while progressively weakening. By the end of plateau-1, the compression ratio approaches unity, indicating an extremely weak shock, consistent with the behavior at the end of the initial rising phase evolution.

\begin{figure}
  \centering
    \includegraphics[angle=0,width=8.8cm,keepaspectratio=true]{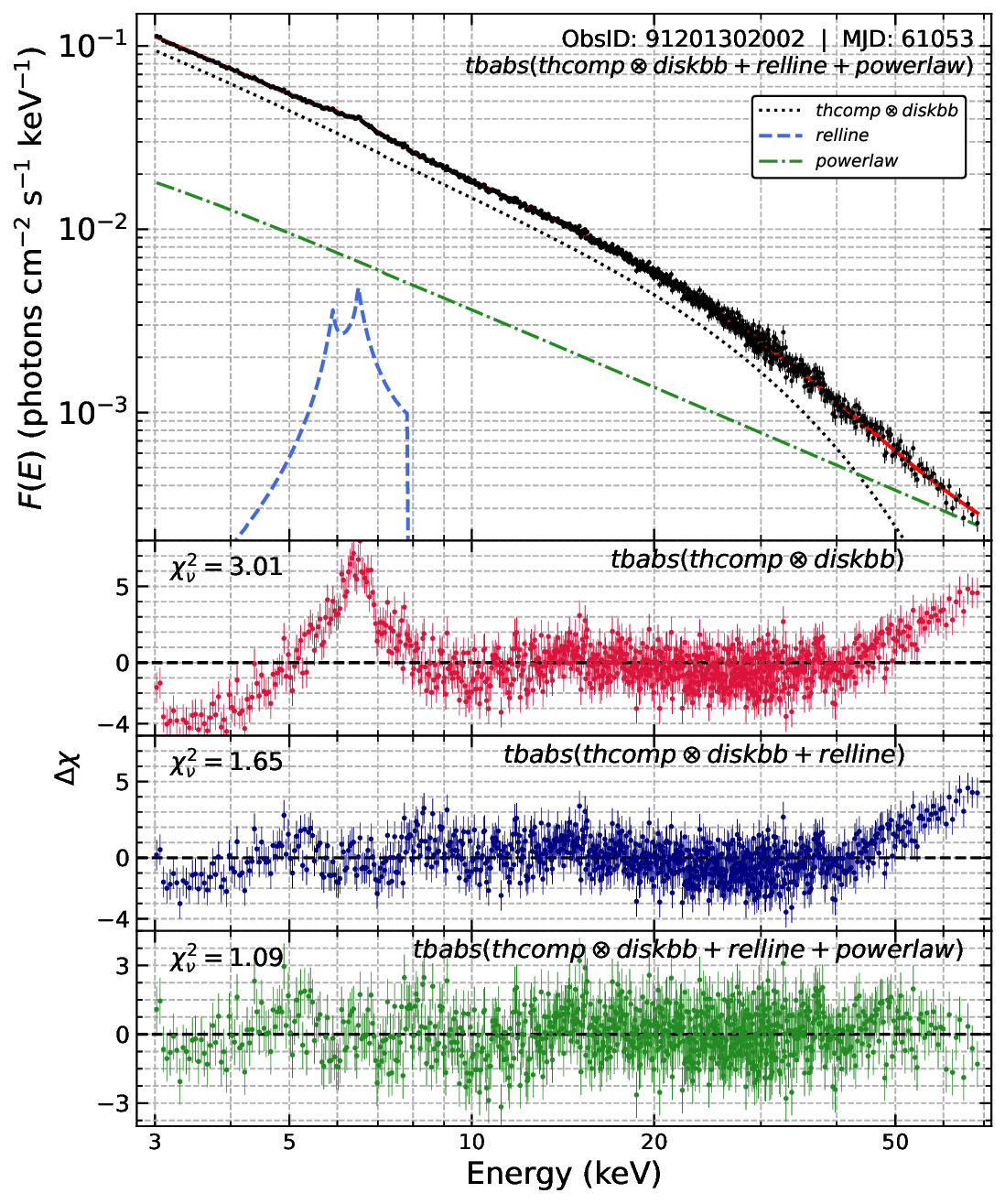}
	\caption{The \textit{NuSTAR}/FPMA spectrum of GS~1354$-$64 obtained on 2026 January 13 (MJD~61053.60; ObsID 91201302002) in the $3$--$70$~keV band. The residual panels show fits with the three spectral models described in the text. The basic model, M1 ($tbabs(thcomp\otimes diskbb)$), leaves significant residuals due to a broad Fe K$_\alpha$ emission line, a reflection hump in the $10$--$30$~keV band, and excess emission above $\sim50$~keV. Adding the \texttt{relline} and \texttt{powerlaw} components successfully accounts for these features and significantly improves the fit.
    }
\label{fig_spec_sin}
\end{figure}

\begin{figure}
  \centering
    \includegraphics[angle=0,width=8.8cm,keepaspectratio=true]{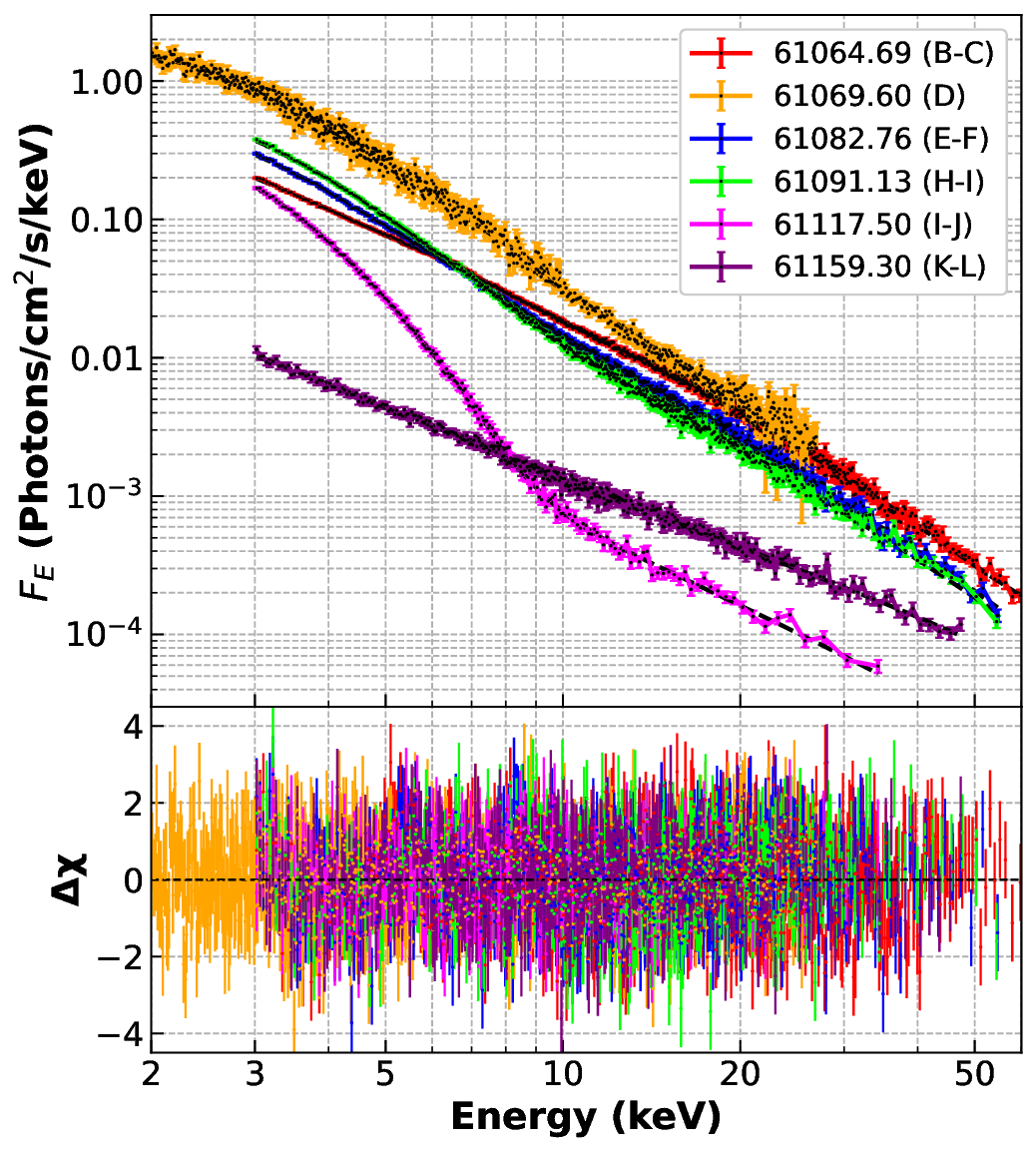}
	\caption{An evolution of model fitted spectra from different observed spectral states of the 2025-2026 outburst are presented. 
	The epochs correspond to MJD 61064.69 is rising HIMS; 61069.60 is peak of the flare-1; 61082.76 and 61091.13 are plateau phases 1 \& 2 
	between flares 1 \& 2 and flares 2 \& 3 respectively; and 61117.50, 61159.30 are declining SIMS and LHS respectively. Note, except 
	peak (MJD 61069.60; \textit{Insight}-HXMT), all other observation are from \textit{NuSTAR}.
	}
\label{fig_spec_evo}
\end{figure}

\subsection{Spectral Analysis}

The evolution of the HRs and LFQPOs provides a broad picture of the source behavior during different phases of the outburst. However, detailed 
spectral modeling of the underlying emission components offers deeper insight into the accretion flow dynamics. The results of the spectral 
analysis are discussed in the following subsections.

\subsubsection{Spectral Modeling}

Broadband spectra from \textit{NuSTAR}/FPMA and the combined LE, ME, and HE instruments of \textit{Insight}-HXMT were used to investigate the spectral 
evolution of GS~1354$-$64 during its 2025--2026 outburst. We initially fitted all spectra with model M1 (see \S2.4). However, the first ten of the eleven 
\textit{NuSTAR} observations (MJD~61053.60--61117.50) showed a prominent broad Fe K$_\alpha$ emission line and reflection features, requiring the addition 
of a relativistic reflection component (\texttt{relline}), defined as model M2. The first three observations (MJD~61053.60--61064.69) also exhibited a 
significant excess above $\sim50$~keV that could not be modelled by M2 alone; therefore, an additional \texttt{powerlaw} component was included, 
defining model M3.

The \textit{NuSTAR}/FPMA spectra are fitted from $3$~keV to SNR limited upper energy range (typically $40$--$70$~keV). Likewise, the combined LE+ME+HE (or LE+ME) spectra from \textit{Insight}-HXMT are fitted from $2$~keV to the highest energy with adequate SNR (typically $20$--$40$~keV, except for the observation on MJD~61058.32, which extends up to $60$~keV). 

Figure~\ref{fig_spec_sin} shows the spectral fit to the first \textit{NuSTAR} observation (ObsID 91201302002; MJD~61053.60) using all three models 
(M1, M2 and M3). In the initial rising LHS, this spectrum exhibits a broad Fe K$_\alpha$ emission line and excess hard X-ray emission above 
$\sim50$~keV. Model M1 adequately describes the continuum below $\sim50$~keV, comprising thermal disk emission and Comptonized photons from the 
`hot' corona. Adding the relativistic reflection component \texttt{relline} (model M2) significantly improves the fit by accounting for the broad 
Fe line and reflection features, reducing the residuals and $\chi_\nu^2$. However, M2 fails to reproduce the emission above $\sim50$~keV. Including 
an additional \texttt{powerlaw} component with $\Gamma_{\rm th} = 1.67\pm0.07$ (model M3) successfully models this high-energy excess and provides the best fit with $\chi_\nu^2 = 1.09$.

For this observation, the \texttt{diskbb} component contributes negligibly, leaving its parameters poorly constrained and preventing reliable 
uncertainty estimates. To test whether this component is statistically required, we replaced the multiplicative Comptonization model \texttt{thcomp} 
with its additive counterpart, \texttt{nthcomp}, and found that the spectrum could be fitted satisfactorily without \texttt{diskbb}. However, to 
ensure consistency across all observations, we retained the \texttt{diskbb}$\otimes$\texttt{thcomp} model and fixed the disk temperature $T_{\rm in}$ 
at its minimum allowed value of $0.25$~keV.

Figure~\ref{fig_spec_evo} presents six representative spectra illustrating the spectral evolution throughout the outburst. The \textit{NuSTAR} 
observation 91201302006 (MJD~61064.69; rising HIMS; B--C) is fitted with model M3. Observations 91202311002 (MJD~61082.76; plateau-1 HIMS; E--F), 
91202311006 (MJD~61091.13; plateau-2 SIMS; H--I), 91202311012 (MJD~61117.50; declining SIMS after the third flare; I--J) and 91202311014 (MJD~61159.30; declining LHS; K--L) are fitted with 
model M2. The remaining \textit{Insight}-HXMT spectrum, corresponding to observation P081404800603 (MJD~61069.60; peak of the first X-ray flare; point~D), is adequately described by the M1 model.

\subsubsection{Spectral Properties}

In Figure~\ref{fig_spec_para}, we show the evolution of the best-fit spectral parameters ($T_{\rm in}$ and $\Gamma_{\rm th}$), model-component fluxes ($F_{dbb}$ and $F_{nth}$), bolometric flux ($F_{bol}$), and QPO frequency ($\nu_{\rm QPO}$). Their evolution closely follows the spectral state transitions described in \S3.1. Except for one observation during the rising HIMS and four during the declining SIMS (yellow-shaded region), where type-B QPOs are detected (red points in Figure~\ref{fig_spec_para}(d)), all QPOs are of type-C. Throughout the 2025--2026 outburst, the disk temperature varied from $0.25$ to $1.04$~keV, while the Comptonized component evolved significantly, with $\Gamma_{\rm th}=1.69$--$2.66$, $kT_e\sim8$--$54$~keV, and $f_{\rm cov}\sim0.02$--$0.97$.

The low photon indices ($\Gamma_{\rm th} < 1.7$), 
and QPO frequencies ($\nu_{\rm QPO} < 0.5$~Hz) during the early rising and late declining phases confirm the source was in the `canonical' LHS 
at both onset and end of the 2025--2026 outburst of GS~1354$-$64.
Following the initial LHS, both $\Gamma_{\rm th}$ and $\nu_{\rm QPO}$ increase monotonically, reaching point C, where the source transits from the rising HIMS 
to the SIMS. The \texttt{diskbb} flux ($F_{\rm dbb}$) shows a similar increase, implying enhanced cooling of the Comptonizing region. This increased 
cooling likely contracts the corona and shifts the shock location inward, resulting in a higher QPO frequency. Throughout the rising harder states 
(A--C), the disk temperature ($T_{\rm in}$) remains nearly constant ($\sim 0.6$~keV), while the bolometric and \texttt{thcomp} fluxes vary only modestly.

The spectral properties evolve rapidly during the first X-ray flare (C--D--E). Along the rising branch (C--D), all flux components increase substantially, 
while the disk temperature ($T_{\rm in}$) rises from $0.63$ to $0.98$~keV and the photon index ($\Gamma_{\rm th}$) from $2.25$ to $2.66$, showing the softest 
spectral nature at the flare peak (point D). No significant QPOs are detected during this phase. Along the declining branch (D--E), the fluxes, $T_{\rm in}$, 
and $\Gamma_{\rm th}$ decrease, indicating progressive spectral hardening. This evolution suggests that the source remained in the SIMS throughout the first flare, 
except for a brief ($\sim1$ day) interval near the peak, when it reached the HSS.


Following the first X-ray flare, the non-thermal \texttt{thcomp} flux ($F_{\rm nth}$) declines steadily until 2026 March 7 (MJD~61106.50). In contrast, 
the bolometric ($F_{\rm bol}$) and \texttt{diskbb} ($F_{\rm dbb}$) fluxes remain nearly constant until point F (2026 February 14; MJD~61085.11), after 
which they increase overall with pronounced short-timescale fluctuations. Excluding the initial $\sim1.5$ days, the interval E--F defines plateau-1, 
during which $T_{\rm in}$, $\Gamma_{\rm th}$, and the type-C QPO frequency ($\nu_{\rm QPO}$) increase gradually. This evolution suggests enhanced cooling of 
the Comptonizing region, driving the shock inward and strengthening the thermal disk emission relative to the non-thermal component. Accordingly, 
plateau-1 is classified as the HIMS. Although its overall evolution resembles the rising HIMS rather than the canonical declining HIMS, the initial 
$\sim1.5$ days are consistent with a typical declining HIMS, marked by decreasing QPO frequency, $F_{\rm dbb}$, $T_{\rm in}$, and $\Gamma_{\rm th}$, indicating 
spectral hardening following the peak of the first flare.

The interval F--G--H (MJD~61085.11--61091.51) corresponds to the second X-ray flare, peaking at point G (MJD~61087.51). At the flare peak, $F_{\rm bol}$, 
$F_{\rm dbb}$, and $\Gamma_{\rm th}$ reach local maxima, indicating a brief transition to the HSS. A similar softening occurs at the peak of the third X-ray flare 
(point I; MJD~61100.51). Excluding these short ($\sim1$ day) HSS episodes, the interval F--J (MJD~61084.51--61123.06) is classified as the declining SIMS, 
during which four type-B LFQPOs are detected. A crossover between $F_{\rm dbb}$ and $F_{\rm nth}$ also occurs at point F, where the thermal disk emission 
becomes dominant over the Comptonized component, further supporting the transition from plateau-1 (HIMS) to the SIMS.

The interval H--I, between the end of the second flare and the peak of the third flare, defines plateau-2. This phase is marked by strong variability, 
with pronounced fluctuations in $F_{\rm bol}$, $F_{\rm dbb}$, $T_{\rm in}$, and $\Gamma_{\rm th}$ indicating rapid changes in the accretion flow geometry and radiative properties.

After the third X-ray flare, the source exhibited the characteristics of a `canonical' declining SIMS. Since \textit{Insight-HXMT} lacked good SNR spectra 
above $7$~keV after the March 7 (MJD~61106.91), we used two archival \textit{NuSTAR} observations to investigate subsequent evolution. In the first 
observation, obtained on 2026 March 18 (MJD~61117.50), spectral component fluxes and the photon index decrease significantly, while the disk temperature remains nearly unchanged relative to the post-flare observations. Consistent with the classification in \S3.1, this observation is identified as the declining SIMS. No significant QPO is detected.

In the next available \textit{NuSTAR} observation, obtained on 2026 April 29 (MJD~61159.30), all flux components, and $\Gamma_{\rm th}$ decrease 
further, indicating continued spectral hardening and confirming the source's return to the LHS. We could not investigate the declining HIMS (J--K) spectrally because no archival observations are available for this phase.

Furthermore, we investigated the evolution of the additional relativistic \texttt{relline} component parameters in the M2 and M3 models, which account for the Fe K$_{\alpha}$ emission line and associated reflection features in the high-resolution \textit{NuSTAR} spectra. The line energy ($E_{\rm line}$) varies from $6.18$ to $6.72$~keV (see Fig.~\ref{fig_spec_evo}d), while the emissivity index ($q^{\star}$) ranges from $\sim1.85$ to $2.42$ throughout the outburst (see Table~\ref{table_master}). In the rising phase (LHS and HIMS; MJD~61053.60--61064.69), $E_{\rm line}$ remains nearly constant at $\sim6.18$--$6.24$~keV. In plateau-1 (HIMS) and the decay of the first X-ray flare (D--E), it increases to $\sim6.29$--$6.42$~keV. At point F, the source enters the SIMS, and $E_{\rm line}$ rises further to $\sim6.63$--$6.72$~keV during plateau-2 (MJD~61085.17--61098.42). In the declining phase (SIMS; MJD~61117.50), it decreases to $\sim6.42$~keV, and subsequently returns to its lower value of $\sim6.18$~keV in the LHS (MJD~61159.30). The \texttt{relline} flux ($F_{\rm RL}$) peaks at $\sim0.07$--$0.08\times10^{-9}$~erg~cm$^{-2}$~s$^{-1}$ around the first X-ray flare (MJD~61070.25--61070.48) and reduces to its minimum of $\sim0.004\times10^{-9}$~erg~cm$^{-2}$~s$^{-1}$ in the declining LHS (MJD~61159.30).

\begin{figure}
  \centering
    \includegraphics[angle=0,width=8.8cm,keepaspectratio=true]{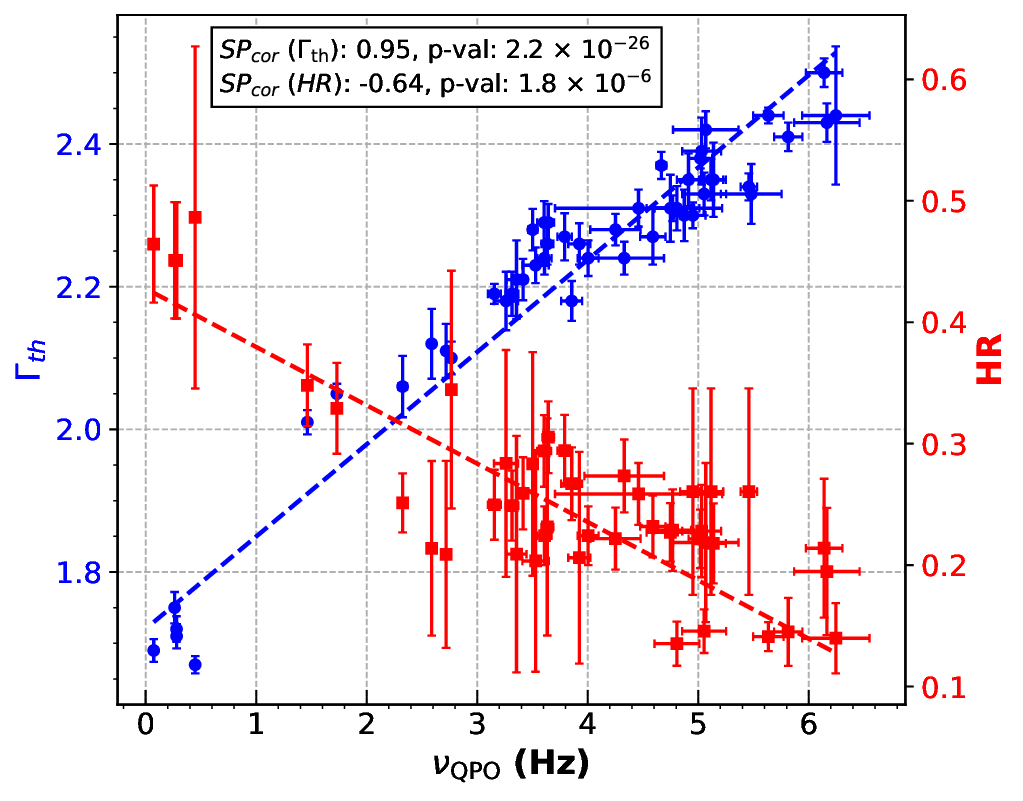}
	\caption{The variation of QPO frequency ($\nu_{QPO}$) with photon index ($\Gamma_{\rm th}$; blue points), and HR (red points), defined 
	as the ratio of MAXI/GSC hard ($6-20$~keV) to soft ($2-6$~keV) fluxes are shown. Linear fit of the correlations between above 
	two sets of parameters are shown by the blue and red dashed-lines.
	}
\label{fig_corr}
\end{figure}

\subsection{Spectro-Temporal Correlation}

The correlation between the timing and spectral properties of BH-XRBs provides valuable insights into the evolution of the accretion flow and Comptonizing corona. In Figure~\ref{fig_corr}, we examine the relationships among the QPO frequency ($\nu_{\rm QPO}$), hardness ratio (HR), and photon index ($\Gamma_{\rm th}$). The HR traces the relative contributions of hard and soft emission from different accretion components, while $\Gamma_{\rm th}$ characterizes the spectral slope and the physical properties of the Comptonizing corona.

The Spearman rank correlation analysis reveals a strong positive correlation between $\nu_{\rm QPO}$ and $\Gamma_{\rm th}$ (SR${\rm cor}=0.95$; p-value 
$=2.2\times10^{-28}$) and a moderately strong negative correlation between $\nu_{\rm QPO}$ and HR (SR${\rm cor}=-0.64$; p-value $=1.8\times10^{-6}$). 
Both correlations are well described by a linear relation, $y=mx+c$. The best-fit $\nu_{\rm QPO}$--$\Gamma_{\rm th}$ relation has $m=0.13\pm0.002$ and 
$c=1.72\pm0.007$, while the $\nu_{\rm QPO}$--HR relation gives $m=-0.05\pm0.004$ and $c=0.43\pm0.02$. These results show that the spectrum softens 
and the hardness ratio decreases with increasing QPO frequency, consistent with enhanced cooling of the Comptonizing region, inward propagation 
of the shock, and the resulting increase in $\nu_{\rm QPO}$.

\section{Discussion and Concluding Remarks}

The renewed activity of the Galactic transient BH-XRB GS~1354$-$64 during its 2025--2026 outburst is studied using on-demand or publicly 
available observations from \textit{MAXI}/GSC, \textit{Insight}-HXMT, and \textit{NuSTAR}. The presence of three short-duration, intense 
X-ray flares makes this outburst complex and challenging to interpret. In this work, we perform a comprehensive spectro-temporal analysis 
to investigate the accretion flow dynamics and the underlying physical processes governing this unusual outburst, which exhibits 
a `non-canonical' evolution of spectral states.
Similar complex, multi-peaked outbursts are not unprecedented and have occasionally been observed in a few transient BH-XRBs (e.g., 1996--1997 and 2005 outbursts of GRO~J1655$-$40, \citealt{Sobczak99, D08}; 1998--1999 outburst of XTE~J1550$-$40, \citealt{Sobczak00}; 1999 outburst of XTE~J1859+226, \citealt{Brocksopp2002}; 2003 outburst of H~1743$-$322, \citealt{McClintock09}).

\textit{MAXI}/GSC monitored the entire outburst with an approximately daily cadence, while \textit{Insight}-HXMT observed the source frequently 
but only over the major activity phase of the outburst. The evolution of the outburst profile in different energy bands and the hardness ratios 
provides an overall picture of nature of the source's characteristics. In the \textit{MAXI}/GSC data, soft X-ray emission dominates the 
intermediate phase, whereas the onset and late phases show comparable soft and hard X-ray fluxes, characteristic of harder spectral states. 
In contrast, the \textit{Insight}-HXMT data are dominated by the broader $6$--$50$~keV hard X-ray emission throughout its observational coverage.

Based on the evolution of the soft and hard X-ray intensities and the hardness ratios from both observatories, we classified the outburst into 
distinct spectral states. The points A--L denote the onset, end, and major transition states. Owing to the presence of three short-duration X-ray 
flares, the source follows an  unusual' sequence of spectral states. Each flare is associated with a pronounced local minimum in the hardness ratio, 
indicating a temporary dominance of soft X-ray emission. The peak intensities of the three flares decrease progressively, reaching approximately 
$1.4$, $0.8$, and $0.5$ Crab, respectively. Although multi-peaked outbursts are common in transient BH-XRBs, only a few systems, such as GRS~1739-278 \citep{Borozdin98}, 4U~1630$-$472 \citep{Baby20,Chatterjee22} and MAXI~J1910$-$057 \citep{Nath23} have shown similarly non-conventional state transitions and temporal evolution.

The HIDs in Figure~\ref{fig_HID} trace the evolution of the hard and soft X-ray emission relative to the total intensity. The \textit{MAXI}/GSC HID1 broadly follows the canonical `q'-shaped track of a complete (type-I) BH-XRB outburst, although its inset highlights the unusually complex evolution of the 2025--2026 outburst of GS~1354$-$64. The \textit{Insight}-HXMT HID2 does not show a complete `q'-track because observations of the initial and late hard states are unavailable. Both HIDs exhibit local spectral softening at the peaks of the three X-ray flares (points D, G, and I). During each flare, the hardness ratio follows a forward-to-backward loop with intensity, indicating spectral softening followed by re-hardening, consistent with the light-curve and HR evolution shown in Figure~\ref{fig_lc}.

As in canonical transient BH-XRB outbursts, GS~1354$-$64 exhibits a monotonic increase in the type-C QPO frequency during the rising pre-flare phase (A--C; Phase-1). Following the first X-ray flare, the QPO reappears at point E and decreases rapidly (Phase-2) before entering a second, longer Phase-3 of increasing frequency (Figure~\ref{fig_spec_para}(e)). While Paper-I explained the first two phases using the POS model \citep{C08,C15}, here we extend the analysis to the third phase and revise the initial fit by including the newly detected 5.76~Hz type-B QPO immediately preceding the first flare. The updated POS model indicates inward shock propagation from $\sim1230$ to $\sim166~r_s$ during Phase-1, outward motion from $\sim151$ to $\sim228~r_s$ during Phase-2, and a slow inward drift from $\sim228$ to $\sim166~r_s$ during Phase-3. These motions reproduce the observed QPO evolution and imply that all three phases either begin or end within a narrow shock location range of $\sim151$--$167~r_s$, suggesting a preferred transition radius for changes in the accretion flow geometry.
As an alternative, the Lense--Thirring (LT) precession model \citep{Stella98,Ingram09,Ingram16} can also be invoked, in which the inner hot flow undergoes global precession, with the resulting QPO frequency depending on the radial extent and structure of the precessing flow. Thus, the two models invoke different physical mechanisms for QPO production: in the POS model, QPOs are associated with oscillations of the post-shock region, whereas in the LT precession model, they arise from the global precession of the inner hot flow.

The reversal from outward to inward shock propagation after Phase-2 is unusual and has previously been reported only in XTE~J1859+226 \citep{N18} and Swift~J1727.8$-$1613 \citep{D25}. We propose that a temporary increase in viscosity at the pile-up radius ($X_p$) during plateau-1 triggered a renewed supply of Keplerian matter. The resulting enhanced Compton cooling halted the outburst decay, drove the shock inward, and likely produced the short-duration X-ray flares. A similar mechanism has been invoked to explain the multi-peaked outbursts of GX~339$-$4 through viscous instabilities induced by irradiation that episodically enhance the mass accretion rate \citep{Aneesha19}. Such events may also produce enhanced optical, and occasionally infrared emission as observed in XTE~J1859+226 \citep{Kimura19} and MAXI~J1820+070 \citep{Russell19}.

Broadband spectral analysis of the 2025--2026 outburst constrains the accretion flow evolution of GS~1354$-$64. The \textit{Insight}-HXMT and \textit{NuSTAR} spectra were fitted with three models (M1, M2 and M3), where \texttt{diskbb} and \texttt{thcomp} represent the thermal disk and Comptonized corona, respectively. All \textit{Insight}-HXMT spectra are well fitted by M1, whereas all eleven \textit{NuSTAR} observations show relativistic reflection features, modeled with an additional \texttt{relline} component (M2) in the last eight observations. The first three rising-phase spectra require an extra \texttt{powerlaw} component ($\Gamma_{\rm PL}\sim1.2$--$1.6$; M3) to explain excess emission above $\sim50$~keV, whose origin remains uncertain.
One possibility is a double-corona geometry \citep{Zdziarski21} observed during the 
hard state of MAXI~J1820+70, although its physical viability is still debated. Alternatively, the excess may arise from reprocessing of 
reflected emission in the corona or at the jet base (via synchrotron cooling processes), consistent with the strong radio brightening 
(see Figure~\ref{fig_lc}(a)) observed during this outburst \citep{Salvaggio26a,Salvaggio26b,Salvaggio26c}. A detailed analysis of the 
reflection-dominated \textit{NuSTAR} spectra using self-consistent physical models will be presented in a forthcoming paper \citep{Bhowmick26}.

The evolution of the spectral parameters ($T_{\rm in}$ and $\Gamma_{\rm th}$), model-component fluxes ($F_{\rm dbb}$, $F_{\rm nth}$, and $F_{\rm bol}$), 
and QPO frequency provides a coherent picture of the spectral state evolution inferred from the outburst profiles and hardness ratios 
(Figure~\ref{fig_lc}). During the initial rising phase (A--C), $\nu_{\rm QPO}$, $\Gamma_{\rm th}$, and $F_{\rm dbb}$ increase together, indicating 
enhanced Compton cooling, a shrinking corona, and an inward-moving shock. Since the type-C QPO frequency scales inversely with the shock 
location ($\nu_{\rm QPO}\propto X_s^{-3/2}$; \citealt{C08}), this interval corresponds to the rising hard states, with A--B in the LHS 
and B--C in the HIMS.

During the first X-ray flare (C--D--E), the spectral parameters and fluxes rise rapidly and then decline, likely due to a temporary increase 
in the supply of Keplerian matter from the pile-up radius ($X_p$; \citealt{C19}). Similar, though weaker, behavior is observed during the 
second and third flares. The absence of QPOs throughout the flare intervals indicates softer spectral states, with the rising and declining 
branches identified as the SIMS and the flare peaks as the HSS. Consistently, all three flare peaks are marked by local maxima in the 
spectral parameters ($T_{\rm in}$ and $\Gamma_{\rm th}$) and fluxes (Figure~\ref{fig_spec_para}).

During plateau-1 (E--F), the first $\sim1.5$ days are dominated by the continued decline of the spectral parameters, model-component fluxes, and $\nu_{\rm QPO}$ following the decay of the first flare. Subsequently, these quantities gradually reverse their trends, except for the non-thermal flux ($F_{\rm nth}$), which decreases steadily until the end of the outburst. The dominance of the non-thermal emission, together with the presence of evolving type-C QPOs, indicates that the source remains in the HIMS throughout plateau-1. At point F, the thermal disk flux exceeds the non-thermal flux for the first time, coincident with the first detection of a type-B QPO, marking the transition to the SIMS. The thermal component remains dominant over the next $\sim40$ days, during which four type-B QPOs are detected.

Plateau-2 (G--I) shows a more complex evolution, with irregular variations in the disk flux, $T_{\rm in}$, and $\Gamma_{\rm th}$, likely driven by a variable supply of high-viscosity Keplerian matter from the pile-up radius. A detailed investigation of the possible connection between these variations and jet activity is beyond the scope of this paper. About $\sim20$ days after the third X-ray flare, the source resumes the `canonical' declining evolution of transient BH-XRBs. The spectral parameters and model-component fluxes gradually decrease, and by the final \textit{NuSTAR} observation the source has returned to the LHS, as indicated by the low QPO frequency, weaker fluxes, lower disk temperature, and harder spectrum. The reflection component is also weakest at this phase, as reflected by the minimum \texttt{relline} flux.

The correlations between the temporal ($\nu_{\rm QPO}$ and HR) and spectral ($\Gamma_{\rm th}$) parameters provide important insights into the evolution of the accretion flow and corona (see Figure~\ref{fig_corr}). We find a strong positive correlation between $\nu_{\rm QPO}$ and $\Gamma_{\rm th}$, indicating that the QPO frequency increases as the spectrum softens. Conversely, $\nu_{\rm QPO}$ is moderately anti-correlated with the hardness ratio, implying that higher QPO frequencies are associated with a larger contribution from soft X-ray emission. In the shock oscillation framework, type-C QPOs arise from resonance oscillations of the `hot' Comptonizing corona, with $\nu_{\rm QPO}$ inversely proportional to the shock location, which defines the outer boundary of the corona. These correlations therefore suggest that a softer spectrum (higher $\Gamma_{\rm th}$ or lower HR) corresponds to a more compact corona, whereas a harder spectrum (lower $\Gamma_{\rm th}$ or higher HR) indicates a larger corona.

A brief summary of findings in this {\it paper} is as follows:

\begin{enumerate}[i)]

    \item During the 2025--2026 outburst, GS~1354$-$64 exhibited a complex outburst profile of $\sim 126$~days, where an unconventional 
    sequence of spectral states, primarily due to the presence of three short-duration ($\sim 2, 5, 7$~days) X-ray flares.  These episodic 
    enhancements in the supply of Keplerian matter not only produced sudden increases in the soft X-ray emission but also delayed the 
    decay of the outburst.
    
    \item In the $2$--$20$~keV band, the \textit{MAXI}/GSC light curves show that the soft X-ray flux dominates over the hard X-ray flux 
    during the middle phase of the outburst (lasting $\sim 90$~days). In contrast, during the initial rising and late declining phases, 
    the soft and hard X-ray fluxes exhibit comparable variability. However, the broadband hard X-ray emission observed by \textit{Insight}-HXMT 
    remains dominant throughout the entire outburst.
    
    \item The three local dips in the HRs during the X-ray flares indicate transient spectral softening toward the HSS, driven by 
    an enhanced soft (thermal) X-ray flux originating from the pre-shock accretion disk.

    \item The HID of \textit{MAXI}/GSC generally follows the canonical `q'-shaped track. However, during the flare phases 
    it exhibits complex forward-to-backward evolutionary loops. This behavior is likely driven by 
    episodic increases in the viscous Keplerian accretion rate, caused by the release of accumulated matter from the pile-up 
    (or accumulation) radius.

    \item A three-phase monotonic evolution of the type-C QPO frequency is observed during the harder spectral states, spanning the 
    initial rising epoch (Phase-1) and plateau-1 (Phases 2 and 3). The QPO frequency increases from $0.07$ to $5.76$ Hz during Phase-1, 
    decreases from $6.14$ to $3.26$ Hz during Phase-2, and then increases again from $3.26$ to $6.25$ Hz during Phase-3.
    
    \item The POS model successfully reproduces the QPO evolution and traces the dynamics of the oscillating shock, which represents the size of the Comptonizing corona. During the rising phase, the shock moves inward from $1230$ to $166~r_s$ while weakening ($R: 4 \rightarrow 1$). During plateau-1, it first recedes from $151$ to $228~r_s$ before slowly propagating inward to $166~r_s$. Notably, all three phases of the QPO evolution either begin or end within a narrow shock location range of $\sim151$--$166~r_s$, suggesting a preferred location for change in accretion flow geometry.

    \item The evolution of the thermal (disk temperature, $T_{\rm in}\sim0.25$--$1.04$~keV) and non-thermal (photon index, $\Gamma_{\rm th}\sim1.69$--$2.66$; coronal electron temperature, $kT_e\sim8$--$54$~keV; covering fraction, $f_{\rm cov}\sim0.02$--$0.97$) spectral parameters, together with the model-component and bolometric fluxes ($F_{\rm dbb}$, $F_{\rm nth}$, $F_{\rm DL}$, and $F_{\rm bol}$) provides a comprehensive view of the accretion flow properties of the source. The complex behavior of the 2025--2026 outburst is consistently explained by the coupled evolution of its spectral and timing properties.

    \item The broad Fe~K$_{\alpha}$ emission line and the associated relativistic reflection features observed in the \textit{NuSTAR} 
    spectra are well described by the \texttt{relline} model. The strength of the reflection component weakens as the source 
    evolves toward the declining LHS.
    
    \item The excess hard X-ray emission above $\sim 50$~keV detected in the \textit{NuSTAR} spectra during the initial rising hard spectral 
    state requires an additional \texttt{powerlaw} component with $\Gamma_{\rm PL} \sim 1.2-1.6$. Although the physical origin of this excess remains 
    uncertain, it may arise from the reprocessing of reflected photons through synchrotron cooling in a strong magnetic field, possibly 
    near the base of the jet.

    \item Furthermore, the observed correlations between the QPO frequency, photon index, and hardness ratio provide valuable insights 
    into the evolving geometry of the accretion flow and the Comptonizing corona.
        
\end{enumerate}

\section*{Acknowledgements}

Authors are thankful to the anonymous referee for valuable suggestions to improve the quality of the manuscript.
This work made use of on-demand data from \textit{MAXI}, a Japanese Experiment Module on the International Space Station, and 
archival data from {\it Insight}-HXMT, a satellite mission of the China National Space Administration (CNSA) and the Chinese 
Academy of Sciences (CAS), as well as \textit{NuSTAR}, a NASA observatory. The authors sincerely thank the MAXI team (RIKEN, JAXA) 
for providing on-demand data products and the HXMT team for providing the latest CALDB file (v2.08) prior to its 
official release.
DD acknowledges the visiting research grant from National Tsing Hua University, Taiwan (NSTC 114-2811-M-007-084). 
RB and HKC acknowledge support from the National Science and Technology Council (NSTC) of Taiwan under grant NSTC 114-2112-M-007-042.
AN thanks GH, SAG; DD; PDMSA; and the Director, URSC, for encouragement and continuous support in carrying out this research. 
We also thank the instrument teams for processing the data and providing the necessary software tools for the analysis.

\bibliography{references_GS}

@PREAMBLE{
  "\providecommand{\apj}{ApJ}" #
  "\providecommand{\apjl}{ApJL}" #
  "\providecommand{\apjs}{ApJS}" #
  "\providecommand{\aap}{A\&A}" #
  "\providecommand{\aapr}{A\&A~Rev.}" #
  "\providecommand{\mnras}{MNRAS}" #
  "\providecommand{\araa}{ARA\&A}" #
  "\providecommand{\pasj}{PASJ}" #
  "\providecommand{\apss}{Ap\&SS}" #
  "\providecommand{\nar}{New~A~Rev.}" #
  "\providecommand{\astl}{Astron.\ Lett.}"
}

@ARTICLE{Belloni02,
       author = {{Belloni}, Tomaso and {Psaltis}, Dimitrios and {van der Klis}, Michiel},
        title = "{A Unified Description of the Timing Features of Accreting X-Ray Binaries}",
      journal = {\apj},
         year = 2002,
        month = jun,
       volume = {572},
       number = {1},
        pages = {392--406},
          doi = {10.1086/340290},
archivePrefix = {arXiv},
       eprint = {astro-ph/0202213},
 primaryClass = {astro-ph},
       adsurl = {https://ui.adsabs.harvard.edu/abs/2002ApJ...572..392B}
}

@ARTICLE{Belloni05,
       author = {{Belloni}, Tomaso and {Homan}, Jeroen and {Casella}, Piergiorgio and
                 {van der Klis}, Michiel and {Nespoli}, Elisa and
                 {Lewin}, Walter H.~G. and {Miller}, Jon M. and {M{\'e}ndez}, Mariano},
        title = "{The evolution of the timing properties of the black-hole transient
                  GX 339-4 during its 2002/2003 outburst}",
      journal = {\aap},
         year = 2005,
        month = sep,
       volume = {440},
       number = {1},
        pages = {207--222},
          doi = {10.1051/0004-6361:20042457},
archivePrefix = {arXiv},
       eprint = {astro-ph/0504577},
 primaryClass = {astro-ph},
       adsurl = {https://ui.adsabs.harvard.edu/abs/2005A%26A...440..207B}
}

@ARTICLE{Belloni12,
       author = {{Belloni}, Tomaso M. and {Sanna}, Andrea and {M{\'e}ndez}, Mariano},
        title = "{High-frequency quasi-periodic oscillations in black hole binaries}",
      journal = {\mnras},
         year = 2012,
        month = nov,
       volume = {426},
       number = {3},
        pages = {1701--1709},
          doi = {10.1111/j.1365-2966.2012.21634.x},
archivePrefix = {arXiv},
       eprint = {1207.2311},
 primaryClass = {astro-ph.HE},
       adsurl = {https://ui.adsabs.harvard.edu/abs/2012MNRAS.426.1701B}
}

@ARTICLE{Borozdin98,
       author = {{Borozdin}, K.~N. and {Revnivtsev}, M.~G. and {Trudolyubov}, S.~P. and
                 {Aleksandrovich}, N.~L. and {Sunyaev}, R.~A. and {Skinner}, G.~K.},
        title = "{The X-ray Nova GRS 1739-278 near the Galactic Center}",
      journal = {\astl},
         year = 1998,
        month = jul,
       volume = {24},
       number = {4},
        pages = {435--444},
archivePrefix = {arXiv},
       eprint = {astro-ph/9901192},
 primaryClass = {astro-ph},
       adsurl = {https://ui.adsabs.harvard.edu/abs/1998AstL...24..435B}
}

@ARTICLE{CT95,
       author = {{Chakrabarti}, Sandip K. and {Titarchuk}, Lev G.},
        title = "{Spectral Properties of Accretion Disks around Galactic and
                  Extragalactic Black Holes}",
      journal = {\apj},
         year = 1995,
        month = dec,
       volume = {455},
        pages = {623},
          doi = {10.1086/176610},
archivePrefix = {arXiv},
       eprint = {astro-ph/9510005},
 primaryClass = {astro-ph},
       adsurl = {https://ui.adsabs.harvard.edu/abs/1995ApJ...455..623C}
}

@ARTICLE{C08,
       author = {{Chakrabarti}, Sandip K. and {Debnath}, Dipak and
                 {Nandi}, Anuj and {Pal}, P.~S.},
        title = "{Evolution of the quasi-periodic oscillation frequency in GRO J1655-40 --
                  Implications for accretion disk dynamics}",
      journal = {\aap},
         year = 2008,
        month = oct,
       volume = {489},
       number = {3},
        pages = {L41--L44},
          doi = {10.1051/0004-6361:200810136},
archivePrefix = {arXiv},
       eprint = {0809.0876},
 primaryClass = {astro-ph},
       adsurl = {https://ui.adsabs.harvard.edu/abs/2008A%26A...489L..41C}
}

@ARTICLE{C19,
       author = {{Chakrabarti}, Sandip K. and {Debnath}, Dipak and {Nagarkoti}, Shreeram},
        title = "{Delayed outburst of H 1743-322 in 2003 and relation with its other outbursts}",
      journal = {Advances in Space Research},
         year = 2019,
       volume = {63},
        pages = {3749},
archivePrefix = {arXiv},
       eprint = {1902.04833},
 primaryClass = {astro-ph.HE},
       adsurl = {https://ui.adsabs.harvard.edu/abs/2019AdSpR..63.3749C}
}

@ARTICLE{C15,
       author = {{Chakrabarti}, Sandip K. and {Mondal}, Santanu and {Debnath}, Dipak},
        title = "{Resonance condition and low-frequency quasi-periodic oscillations of
                  the outbursting source H1743-322}",
      journal = {\mnras},
         year = 2015,
        month = sep,
       volume = {452},
       number = {4},
        pages = {3451--3456},
          doi = {10.1093/mnras/stv1566},
       adsurl = {https://ui.adsabs.harvard.edu/abs/2015MNRAS.452.3451C}
}

@ARTICLE{Chen97,
       author = {{Chen}, Wan and {Shrader}, C.~R. and {Livio}, Mario},
        title = "{The Properties of X-Ray and Optical Light Curves of X-Ray Novae}",
      journal = {\apj},
         year = 1997,
        month = dec,
       volume = {491},
       number = {1},
        pages = {312--338},
          doi = {10.1086/304921},
archivePrefix = {arXiv},
       eprint = {astro-ph/9707138},
 primaryClass = {astro-ph},
       adsurl = {https://ui.adsabs.harvard.edu/abs/1997ApJ...491..312C}
}

@ARTICLE{D13,
       author = {{Debnath}, Dipak and {Chakrabarti}, Sandip K. and {Nandi}, Anuj},
        title = "{Evolution of the temporal and the spectral properties in 2010 and 2011
                  outbursts of H 1743-322}",
      journal = {Advances in Space Research},
         year = 2013,
        month = dec,
       volume = {52},
       number = {11},
        pages = {2143--2155},
          doi = {10.1016/j.asr.2013.09.006},
archivePrefix = {arXiv},
       eprint = {1309.2564},
 primaryClass = {astro-ph.HE},
       adsurl = {https://ui.adsabs.harvard.edu/abs/2013AdSpR..52.2143D}
}

@ARTICLE{D15,
       author = {{Debnath}, Dipak and {Mondal}, Santanu and {Chakrabarti}, Sandip K.},
        title = "{Characterization of GX 339-4 outburst of 2010-11: analysis by XSPEC
                  using two component advective flow model}",
      journal = {\mnras},
         year = 2015,
        month = feb,
       volume = {447},
       number = {2},
        pages = {1984--1995},
          doi = {10.1093/mnras/stu2588},
archivePrefix = {arXiv},
       eprint = {1306.3745},
 primaryClass = {astro-ph.HE},
       adsurl = {https://ui.adsabs.harvard.edu/abs/2015MNRAS.447.1984D}
}

@ARTICLE{D17,
       author = {{Debnath}, Dipak and {Jana}, Arghajit and {Chakrabarti}, Sandip K. and
                 {Chatterjee}, Debjit and {Mondal}, Santanu},
        title = "{Accretion Flow Properties of Swift J1753.5-0127 during Its 2005 Outburst}",
      journal = {\apj},
         year = 2017,
        month = nov,
       volume = {850},
       number = {1},
          eid = {92},
        pages = {92},
          doi = {10.3847/1538-4357/aa9077},
archivePrefix = {arXiv},
       eprint = {1703.05479},
 primaryClass = {astro-ph.HE},
       adsurl = {https://ui.adsabs.harvard.edu/abs/2017ApJ...850...92D}
}

@ARTICLE{D20,
       author = {{Debnath}, Dipak and {Chatterjee}, Debjit and {Jana}, Arghajit and
                 {Chakrabarti}, Sandip K. and {Chatterjee}, Kaushik},
        title = "{Accretion flow properties of XTE J1118+480 during its 2005 outburst}",
      journal = {Research in Astronomy and Astrophysics},
         year = 2020,
       volume = {20},
       number = {11},
          eid = {175},
        pages = {175},
          doi = {10.1088/1674-4527/20/11/175},
       adsurl = {https://ui.adsabs.harvard.edu/abs/2020RAA....20..175D}
}

@ARTICLE{D21,
       author = {{Debnath}, Dipak and {Chatterjee}, Kaushik and {Chatterjee}, Debjit and
                 {Jana}, Arghajit and {Chakrabarti}, Sandip K.},
        title = "{Jet properties of XTE J1752-223 during its 2009-2010 outburst}",
      journal = {\mnras},
         year = 2021,
       volume = {504},
       number = {3},
        pages = {4242},
          doi = {10.1093/mnras/stab1169},
archivePrefix = {arXiv},
       eprint = {2104.10485},
 primaryClass = {astro-ph.HE},
       adsurl = {https://ui.adsabs.harvard.edu/abs/2021MNRAS.504.4242D}
}

@ARTICLE{D25,
       author = {{Debnath}, Dipak and {Srimani}, Subham and {Chang}, Hsiang-Kuang},
        title = "{Detection of X-Ray Polarization in the Hard State of IGR J17091-3624:
                  Spectropolarimetric Study with IXPE and NuSTAR Data}",
      journal = {\apj},
         year = 2025,
       volume = {989},
       number = {2},
          eid = {165},
        pages = {165},
          doi = {10.3847/1538-4357/adf1e6},
archivePrefix = {arXiv},
       eprint = {2504.16402},
 primaryClass = {astro-ph.HE},
       adsurl = {https://ui.adsabs.harvard.edu/abs/2025ApJ...989..165D}
}

@ARTICLE{D26,
       author = {{Debnath}, Dipak and {Chang}, Hsiang-Kuang and {Srimani}, Subham and {Nandi}, Anuj},
        title = "{Multimission Observations of GS 1354-64 during the 2025--2026 Outburst:
                  First Results}",
      journal = {\apj},
         year = 2026,
        month = jun,
       volume = {1004},
       number = {2},
          eid = {238},
        pages = {238},
          doi = {10.3847/1538-4357/ae6f0c},
archivePrefix = {arXiv},
       eprint = {2605.27415},
 primaryClass = {astro-ph.HE},
       adsurl = {https://ui.adsabs.harvard.edu/abs/2026ApJ..1004..238D}
}

@ARTICLE{D08,
       author = {{Debnath}, D. and {Chakrabarti}, S.~K. and {Nandi}, A. and {Mandal}, S.},
        title = "{Spectral and timing evolution of GRO J1655-40 during its outburst of 2005}",
      journal = {Bulletin of the Astronomical Society of India},
         year = 2008,
       volume = {36},
        pages = {151},
archivePrefix = {arXiv},
       eprint = {0902.3791},
 primaryClass = {astro-ph.HE}
}

@ARTICLE{D10,
       author = {{Debnath}, D. and {Chakrabarti}, S.~K. and {Nandi}, A.},
        title = "{Properties of the propagating shock wave in the accretion flow around
                  GX 339-4 in the 2010 outburst}",
      journal = {\aap},
         year = 2010,
       volume = {520},
          eid = {A98},
        pages = {A98},
          doi = {10.1051/0004-6361/201014990},
archivePrefix = {arXiv},
       eprint = {1009.3351},
 primaryClass = {astro-ph.HE},
       adsurl = {https://ui.adsabs.harvard.edu/abs/2010A%26A...520A..98D}
}

@ARTICLE{Done07,
       author = {{Done}, Chris and {Gierli{\'n}ski}, Marek and {Kubota}, Aya},
        title = "{Modelling the behaviour of accretion flows in X-ray binaries. Everything
                  you always wanted to know about accretion but were afraid to ask}",
      journal = {\aapr},
         year = 2007,
        month = dec,
       volume = {15},
       number = {1},
        pages = {1--66},
          doi = {10.1007/s00159-007-0006-1},
archivePrefix = {arXiv},
       eprint = {0708.0148},
 primaryClass = {astro-ph},
       adsurl = {https://ui.adsabs.harvard.edu/abs/2007A%26ARv..15....1D}
}

@ARTICLE{Fabian89,
       author = {{Fabian}, A.~C. and {Rees}, M.~J. and {Stella}, L. and {White}, N.~E.},
        title = "{X-ray fluorescence from the inner disc in Cygnus X-1}",
      journal = {\mnras},
         year = 1989,
        month = jun,
       volume = {238},
       number = {3},
        pages = {729--736},
          doi = {10.1093/mnras/238.3.729},
       adsurl = {https://ui.adsabs.harvard.edu/abs/1989MNRAS.238..729F}
}

@ARTICLE{Fender04,
       author = {{Fender}, R.~P. and {Belloni}, T.~M. and {Gallo}, E.},
        title = "{Towards a unified model for black hole X-ray binary jets}",
      journal = {\mnras},
         year = 2004,
        month = dec,
       volume = {355},
       number = {4},
        pages = {1105--1118},
          doi = {10.1111/j.1365-2966.2004.08384.x},
archivePrefix = {arXiv},
       eprint = {astro-ph/0409360},
 primaryClass = {astro-ph},
       adsurl = {https://ui.adsabs.harvard.edu/abs/2004MNRAS.355.1105F}
}

@ARTICLE{Fender09,
       author = {{Fender}, R.~P. and {Homan}, J. and {Belloni}, T.~M.},
        title = "{Jets from black hole X-ray binaries: testing, refining and extending
                  empirical models for the coupling to X-rays}",
      journal = {\mnras},
         year = 2009,
        month = jul,
       volume = {396},
       number = {3},
        pages = {1370--1382},
          doi = {10.1111/j.1365-2966.2009.14841.x},
archivePrefix = {arXiv},
       eprint = {0903.5166},
 primaryClass = {astro-ph.HE},
       adsurl = {https://ui.adsabs.harvard.edu/abs/2009MNRAS.396.1370F}
}

@ARTICLE{Garcia10,
       author = {{Garc{\'\i}a}, J. and {Kallman}, T.~R.},
        title = "{X-ray Reflected Spectra from Accretion Disk Models. I. Constant Density
                  Atmospheres}",
      journal = {\apj},
         year = 2010,
        month = aug,
       volume = {718},
       number = {2},
        pages = {695--701},
          doi = {10.1088/0004-637X/718/2/695},
archivePrefix = {arXiv},
       eprint = {1006.0485},
 primaryClass = {astro-ph.HE},
       adsurl = {https://ui.adsabs.harvard.edu/abs/2010ApJ...718..695G}
}

@ARTICLE{Iyer15,
       author = {{Iyer}, N. and {Nandi}, A. and {Mandal}, S.},
        title = "{Determination of Mass of IGR J17091-3624 from ``Spectro-temporal''
                  Variations during Onset-phase of the 2011 Outburst}",
      journal = {\apj},
         year = 2015,
        month = jul,
       volume = {807},
       number = {1},
          eid = {108},
        pages = {108},
          doi = {10.1088/0004-637X/807/1/108},
archivePrefix = {arXiv},
       eprint = {1505.02529},
 primaryClass = {astro-ph.HE},
       adsurl = {https://ui.adsabs.harvard.edu/abs/2015ApJ...807..108I}
}

@ARTICLE{Jana17,
       author = {{Jana}, Arghajit and {Chakrabarti}, Sandip K. and {Debnath}, Dipak},
        title = "{Properties of X-Ray Flux of Jets during the 2005 Outburst of
                  Swift J1753.5-0127 Using the TCAF Solution}",
      journal = {\apj},
         year = 2017,
       volume = {850},
       number = {1},
          eid = {91},
        pages = {91},
          doi = {10.3847/1538-4357/aa88a5},
archivePrefix = {arXiv},
       eprint = {1708.08054},
 primaryClass = {astro-ph.HE},
       adsurl = {https://ui.adsabs.harvard.edu/abs/2017ApJ...850...91J}
}

@ARTICLE{Kocevski03,
       author = {{Kocevski}, Daniel and {Ryde}, Felix and {Liang}, Edison},
        title = "{Search for Relativistic Curvature Effects in Gamma-Ray Burst Pulses}",
      journal = {\apj},
         year = 2003,
        month = oct,
       volume = {596},
       number = {1},
        pages = {389--400},
          doi = {10.1086/377707},
archivePrefix = {arXiv},
       eprint = {astro-ph/0303556},
 primaryClass = {astro-ph},
       adsurl = {https://ui.adsabs.harvard.edu/abs/2003ApJ...596..389K}
}

@ARTICLE{Lasota01,
       author = {{Lasota}, Jean-Pierre},
        title = "{The disc instability model of dwarf novae and low-mass X-ray binary transients}",
      journal = {\nar},
         year = 2001,
        month = jun,
       volume = {45},
       number = {7},
        pages = {449--508},
          doi = {10.1016/S1387-6473(01)00112-9},
archivePrefix = {arXiv},
       eprint = {astro-ph/0102072},
 primaryClass = {astro-ph},
       adsurl = {https://ui.adsabs.harvard.edu/abs/2001NewAR..45..449L}
}

@ARTICLE{Lightman88,
       author = {{Lightman}, Alan P. and {White}, Tim R.},
        title = "{Effects of Cold Matter in Active Galactic Nuclei: A Broad Hump in the
                  X-Ray Spectra}",
      journal = {\apj},
         year = 1988,
        month = dec,
       volume = {335},
        pages = {57--66},
          doi = {10.1086/166905},
       adsurl = {https://ui.adsabs.harvard.edu/abs/1988ApJ...335...57L}
}

@ARTICLE{N01,
       author = {{Nandi}, A. and {Chakrabarti}, S.~K. and {Vadawale}, S.~V. and {Rao}, A.~R.},
        title = "{Ejection of the inner accretion disk in GRS 1915+105: the magnetic
                  rubber-band effect}",
      journal = {\aap},
         year = 2001,
        month = dec,
       volume = {380},
       number = {1},
        pages = {245--250},
          doi = {10.1051/0004-6361:20011444},
       adsurl = {https://ui.adsabs.harvard.edu/abs/2001A%26A...380..245N}
}

@ARTICLE{N12,
       author = {{Nandi}, A. and {Debnath}, D. and {Mandal}, S. and {Chakrabarti}, S.~K.},
        title = "{Accretion flow dynamics during the evolution of timing and spectral
                  properties of GX 339-4 during its 2010-11 outburst}",
      journal = {\aap},
         year = 2012,
        month = jun,
       volume = {542},
          eid = {A56},
        pages = {A56},
          doi = {10.1051/0004-6361/201117844},
archivePrefix = {arXiv},
       eprint = {1204.5044},
 primaryClass = {astro-ph.HE},
       adsurl = {https://ui.adsabs.harvard.edu/abs/2012A%26A...542A..56N}
}

@ARTICLE{N24,
       author = {{Nandi}, Anuj and {Das}, Santabrata and {Majumder}, Seshadri and
                 {Katoch}, Tilak and {Antia}, H.~M. and {Shah}, Parag},
        title = "{Discovery of evolving low-frequency QPOs in hard X-rays ($\sim$100 keV)
                  observed in black hole Swift J1727.8-1613 with AstroSat}",
      journal = {\mnras},
         year = 2024,
        month = jun,
       volume = {531},
       number = {1},
        pages = {1149--1157},
          doi = {10.1093/mnras/stae1208},
archivePrefix = {arXiv},
       eprint = {2404.17160},
 primaryClass = {astro-ph.HE},
       adsurl = {https://ui.adsabs.harvard.edu/abs/2024MNRAS.531.1149N}
}

@ARTICLE{Radhika16,
       author = {{Radhika}, D. and {Nandi}, A. and {Agrawal}, V.~K. and {Seetha}, S.},
        title = "{`Spectro-temporal' variabilities and possible physical mechanism for
                  jet ejections}",
      journal = {\mnras},
         year = 2016,
       volume = {460},
       number = {4},
        pages = {4403--4416},
          doi = {10.1093/mnras/stw1239},
archivePrefix = {arXiv},
       eprint = {1605.08351},
 primaryClass = {astro-ph.HE},
       adsurl = {https://ui.adsabs.harvard.edu/abs/2016MNRAS.460.4403R}
}

@ARTICLE{RM06,
       author = {{Remillard}, Ronald A. and {McClintock}, Jeffrey E.},
        title = "{X-Ray Properties of Black-Hole Binaries}",
      journal = {\araa},
         year = 2006,
        month = sep,
       volume = {44},
       number = {1},
        pages = {49--92},
          doi = {10.1146/annurev.astro.44.051905.092532},
archivePrefix = {arXiv},
       eprint = {astro-ph/0606352},
 primaryClass = {astro-ph},
       adsurl = {https://ui.adsabs.harvard.edu/abs/2006ARA%26A..44...49R}
}

@ARTICLE{SS73,
       author = {{Shakura}, N.~I. and {Sunyaev}, R.~A.},
        title = "{Black holes in binary systems. Observational appearance.}",
      journal = {\aap},
         year = 1973,
        month = jan,
       volume = {24},
        pages = {337--355},
       adsurl = {https://ui.adsabs.harvard.edu/abs/1973A%26A....24..337S}
}

@ARTICLE{ST80,
       author = {{Sunyaev}, R.~A. and {Titarchuk}, L.~G.},
        title = "{Comptonization of X-rays in plasma clouds. Typical radiation spectra.}",
      journal = {\aap},
         year = 1980,
       volume = {86},
        pages = {121--138},
       adsurl = {https://ui.adsabs.harvard.edu/abs/1980A%26A....86..121S}
}

@ARTICLE{ST85,
       author = {{Sunyaev}, R.~A. and {Titarchuk}, L.~G.},
        title = "{Comptonization of low-frequency radiation in accretion disks: Angular
                  distribution and polarization of hard radiation}",
      journal = {\aap},
         year = 1985,
        month = feb,
       volume = {143},
        pages = {374--388},
       adsurl = {https://ui.adsabs.harvard.edu/abs/1985A%26A...143..374S}
}

@ARTICLE{Tanaka96,
       author = {{Tanaka}, Yasuo and {Shibazaki}, Noriaki},
        title = "{X-ray Novae}",
      journal = {\araa},
         year = 1996,
        month = jan,
       volume = {34},
        pages = {607--644},
          doi = {10.1146/annurev.astro.34.1.607},
       adsurl = {https://ui.adsabs.harvard.edu/abs/1996ARA%26A..34..607T}
}

@ARTICLE{vanderKlis94,
       author = {{van der Klis}, M.},
        title = "{A comparison of the power spectra of Z and atoll sources, pulsars and
                  black hole candidates}",
      journal = {\aap},
         year = 1994,
       volume = {283},
        pages = {469--474},
       adsurl = {https://ui.adsabs.harvard.edu/abs/1994A%26A...283..469V}
}

@INPROCEEDINGS{Zhang14,
       author = {{Zhang}, S. and {Lu}, F.~J. and {Zhang}, S.~N. and {Li}, T.~P.},
        title = "{Introduction to the hard X-ray modulation telescope}",
    booktitle = {Space Telescopes and Instrumentation 2014: Ultraviolet to Gamma Ray},
       series = {Society of Photo-Optical Instrumentation Engineers (SPIE) Conference Series},
         year = 2014,
        month = jul,
       volume = {9144},
          eid = {914421},
        pages = {914421},
          doi = {10.1117/12.2054144},
       adsurl = {https://ui.adsabs.harvard.edu/abs/2014SPIE.9144E..21Z}
}

@ARTICLE{Zhang19,
       author = {{Zhang}, G.-B. and {Bernardini}, F. and {Russell}, D.~M. and
                 {Gelfand}, J.~D. and {Lasota}, J.-P. and {Al Qasim}, A. and
                 {AlMannaei}, A. and {Koljonen}, K.~I.~I. and {Shaw}, A.~W. and
                 {Lewis}, F. and {Tomsick}, J.~A. and {Plotkin}, R.~M. and
                 {Miller-Jones}, J.~C.~A. and {Maitra}, D. and {Homan}, J. and
                 {Charles}, P.~A. and {Kobel}, P. and {Perez}, D. and {Doran}, R.},
        title = "{Bright Mini-outburst Ends the 12 yr Long Activity of the Black Hole
                  Candidate Swift J1753.5-0127}",
      journal = {\apj},
         year = 2019,
        month = may,
       volume = {876},
       number = {1},
          eid = {5},
        pages = {5},
          doi = {10.3847/1538-4357/ab12dd},
archivePrefix = {arXiv},
       eprint = {1903.09455},
 primaryClass = {astro-ph.HE},
       adsurl = {https://ui.adsabs.harvard.edu/abs/2019ApJ...876....5Z}
}

@ARTICLE{Zhang20,
       author = {{Zhang}, Shuang-Nan and {Li}, Tipei and {Lu}, Fangjun and {Song}, Liming and
                 {Xu}, Yupeng and {Liu}, Congzhan and others},
        title = "{Overview to the Hard X-ray Modulation Telescope (Insight-HXMT) Satellite}",
      journal = {Science China Physics, Mechanics, and Astronomy},
         year = 2020,
        month = apr,
       volume = {63},
       number = {4},
          eid = {249502},
        pages = {249502},
          doi = {10.1007/s11433-019-1432-6},
archivePrefix = {arXiv},
       eprint = {1910.09613},
 primaryClass = {astro-ph.IM},
       adsurl = {https://ui.adsabs.harvard.edu/abs/2020SCPMA..6349502Z}
}

@ARTICLE{Choudhury25,
       author = {{Das Choudhury}, Sreetama and {Bhuvana}, G.~R. and {Das}, Santabrata and {Nandi}, Anuj},
        title = "{Revisiting disc-jet coupling in black hole X-ray binaries: on the nature
                  of disc dynamics and jet velocity}",
      journal = {\mnras},
         year = 2025,
       volume = {541},
       number = {4},
        pages = {2934--2954},
          doi = {10.1093/mnras/staf1107},
archivePrefix = {arXiv},
       eprint = {2507.03644},
 primaryClass = {astro-ph.HE},
       adsurl = {https://ui.adsabs.harvard.edu/abs/2025MNRAS.541.2934D}
}

@ARTICLE{Aneesha19,
       author = {{Aneesha}, U. and {Mandal}, S. and {Sreehari}, H.},
        title = "{Study of the long-term evolution of the accretion dynamics of GX 339-4}",
      journal = {\mnras},
         year = 2019,
       volume = {486},
       number = {2},
        pages = {2705--2720},
          doi = {10.1093/mnras/stz1000},
archivePrefix = {arXiv},
       eprint = {1905.04553},
 primaryClass = {astro-ph.HE},
       adsurl = {https://ui.adsabs.harvard.edu/abs/2019MNRAS.486.2705A}
}

@ARTICLE{Baby20,
       author = {{Baby}, Blessy E. and {Agrawal}, V.~K. and {Ramadevi}, M.~C. and
                 {Katoch}, Tilak and {Antia}, H.~M. and {Mandal}, Samir and {Nandi}, Anuj},
        title = "{AstroSat and MAXI view of the black hole binary 4U 1630-472 during 2016
                  and 2018 outbursts}",
      journal = {\mnras},
         year = 2020,
       volume = {497},
        pages = {1197--1211},
          doi = {10.1093/mnras/staa1965},
       adsurl = {https://ui.adsabs.harvard.edu/abs/2020MNRAS.497.1197B}
}

@ARTICLE{Casares04,
       author = {{Casares}, J. and {Zurita}, C. and {Shahbaz}, T. and {Charles}, P.~A. and {Fender}, R.~P.},
        title = "{Evidence for a Black Hole in the X-Ray Transient GS 1354-64 (=BW Circini)}",
      journal = {\apjl},
         year = 2004,
        month = oct,
       volume = {613},
       number = {2},
        pages = {L133--L136},
          doi = {10.1086/425145},
archivePrefix = {arXiv},
       eprint = {astro-ph/0408331},
 primaryClass = {astro-ph},
       adsurl = {https://ui.adsabs.harvard.edu/abs/2004ApJ...613L.133C}
}

@ARTICLE{Casares09,
       author = {{Casares}, J. and {Orosz}, J.~A. and {Zurita}, C. and {Shahbaz}, T. and
                 {Corral-Santana}, J.~M. and {McClintock}, J.~E. and {Garcia}, M.~R. and
                 {Mart{\'\i}nez-Pais}, I.~G. and {Charles}, P.~A. and {Fender}, R.~P. and
                 {Remillard}, R.~A.},
        title = "{Refined Orbital Solution and Quiescent Variability in the Black Hole
                  Transient GS 1354-64 (=BW Cir)}",
      journal = {\apjs},
         year = 2009,
        month = mar,
       volume = {181},
       number = {1},
        pages = {238--243},
          doi = {10.1088/0067-0049/181/1/238},
       adsurl = {https://ui.adsabs.harvard.edu/abs/2009ApJS..181..238C}
}

@ARTICLE{El-Batal16,
       author = {{El-Batal}, A.~M. and {Miller}, J.~M. and {Reynolds}, M.~T. and {Boggs}, S.~E. and
                 {Christensen}, F.~E. and {Craig}, W.~W. and {F{\"u}rst}, F. and {Hailey}, C.~J. and
                 {Harrison}, F.~A. and {Stern}, D.~K. and {Tomsick}, J. and {Walton}, D.~J. and
                 {Zhang}, W.~W.},
        title = "{NuSTAR Observations of the Black Hole GS 1354-645: Evidence of Rapid
                  Black Hole Spin}",
      journal = {\apjl},
         year = 2016,
        month = jul,
       volume = {826},
       number = {1},
          eid = {L12},
        pages = {L12},
          doi = {10.3847/2041-8205/826/1/L12},
archivePrefix = {arXiv},
       eprint = {1607.00343},
 primaryClass = {astro-ph.HE},
       adsurl = {https://ui.adsabs.harvard.edu/abs/2016ApJ...826L..12E}
}

@ARTICLE{Hannikainen05,
       author = {{Hannikainen}, D.~C. and {Charles}, P.~A. and {van Zyl}, L. and {Kong}, A.~K.~H. and
                 {Homer}, L. and {Hakala}, P. and {Naylor}, T. and {Davies}, M.~B.},
        title = "{The X-ray source population of the globular cluster M15: Chandra
                  high-resolution imaging}",
      journal = {\mnras},
         year = 2005,
       volume = {357},
       number = {1},
        pages = {325--332},
          doi = {10.1111/j.1365-2966.2004.08652.x},
       adsurl = {https://ui.adsabs.harvard.edu/abs/2005MNRAS.357..325H}
}

@ARTICLE{Kitamoto90,
       author = {{Kitamoto}, S. and {Tsunemi}, H. and {Pedersen}, H. and {Ilovaisky}, S.~A. and
                 {van der Klis}, M.},
        title = "{Outburst, Identification, and X-Ray Light Curve of GS 1354-64
                  (= MX 1353-64, Centaurus X-2?)}",
      journal = {\apj},
         year = 1990,
       volume = {361},
        pages = {590},
       adsurl = {https://ui.adsabs.harvard.edu/abs/1990ApJ...361..590K}
}

@ARTICLE{Liu26,
       author = {{Liu}, Honghui and {Kong}, Lingda and {Adegoke}, Oluwashina K. and
                 {Jiang}, Jiachen and {Bambi}, Cosimo and {Fabian}, Andrew C. and
                 {Ingram}, Adam and {Ravi}, Swati and {Steiner}, James F. and
                 {Shui}, Qingcang and {Walton}, Dominic J. and {Xu}, Yerong and
                 {Young}, Andrew J. and {Zhang}, Yuexin and {Zhang}, Zuobin and
                 {Santangelo}, Andrea},
        title = "{A clean broad iron line in GS 1354-64 as seen by XRISM}",
      journal = {arXiv e-prints},
         year = 2026,
        month = mar,
archivePrefix = {arXiv},
       eprint = {2603.06883},
 primaryClass = {astro-ph.HE}
}

@ARTICLE{N18,
       author = {{Nandi}, A. and {Mandal}, S. and {Sreehari}, H. and {Radhika}, D. and {Das}, S. and
                 {Chattopadhyay}, I. and {Iyer}, N. and {Agrawal}, V.~K. and {Aktar}, R.},
        title = "{Accretion flow dynamics during 1999 outburst of XTE J1859+226 -- modeling
                  of broadband spectra and constraining the source mass}",
      journal = {\apss},
         year = 2018,
       volume = {363},
       number = {5},
          eid = {90},
        pages = {90},
          doi = {10.1007/s10509-018-3314-1},
       adsurl = {https://ui.adsabs.harvard.edu/abs/2018Ap%26SS.363...90N}
}

@ARTICLE{Ravi26,
       author = {{Ravi}, Swati and {Marra}, Lorenzo and {Steiner}, James F. and
                 {Mastroserio}, Guglielmo and {Ng}, Mason and {Neilsen}, Joey and
                 {Marshall}, Herman L. and others},
        title = "{Discovery of Strong Energy-Dependent X-ray Polarization in the
                  Intermediate State of GS 1354-64}",
      journal = {arXiv e-prints},
         year = 2026,
        month = mar,
archivePrefix = {arXiv},
       eprint = {2603.03463},
 primaryClass = {astro-ph.HE}
}

@ARTICLE{Sreehari19,
       author = {{Sreehari}, H. and {Iyer}, N. and {Radhika}, D. and {Nandi}, A. and {Mandal}, S.},
        title = "{Constraining the mass of the black hole GX 339-4 using
                  spectro-temporal analysis of multiple outbursts}",
      journal = {Advances in Space Research},
         year = 2019,
       volume = {63},
       number = {3},
        pages = {1374--1386},
          doi = {10.1016/j.asr.2018.10.042},
archivePrefix = {arXiv},
       eprint = {1811.04341},
 primaryClass = {astro-ph.HE},
       adsurl = {https://ui.adsabs.harvard.edu/abs/2019AdSpR..63.1374S}
}

@ARTICLE{Bhowmick21,
       author = {{Bhowmick}, Riya and {Debnath}, Dipak and {Chatterjee}, Kaushik and
                 {Nagarkoti}, Shreeram and {Chakrabarti}, Sandip K. and {Sarkar}, Ritabrata and
                 {Chatterjee}, Debjit and {Jana}, Arghajit},
        title = "{Relation between Quiescence and Outbursting Properties of GX 339-4}",
      journal = {\apj},
         year = 2021,
       volume = {910},
        pages = {138},
        doi = {10.3847/1538-4357/abe134},
       adsurl = {https://ui.adsabs.harvard.edu/abs/2021ApJ...910..138B}
}

@ARTICLE{Chatterjee19,
       author = {{Chatterjee}, Debjit and {Debnath}, Dipak and {Jana}, Arghajit and {Chakrabarti}, Sandip K.},
        title = "{Properties of the black hole candidate XTE J1118+480 with the TCAF
                  solution during its jet activity induced 2000 outburst}",
      journal = {\apss},
         year = 2019,
       volume = {364},
        pages = {14},
archivePrefix = {arXiv},
       eprint = {1803.09438},
 primaryClass = {astro-ph.HE},
       adsurl = {https://ui.adsabs.harvard.edu/abs/2019Ap%26SS.364...14C}
}

@ARTICLE{Chatterjee22,
       author = {{Chatterjee}, Kaushik and {Debnath}, Dipak and {Bhowmick}, Riya and
                 {Nath}, Sujoy Kumar and {Chatterjee}, Debjit},
        title = "{Anomalous nature of outbursts of black hole candidate 4U 1630-472}",
      journal = {\mnras},
         year = 2022,
       volume = {510},
        pages = {1128},
          doi = {10.1093/mnras/stab3570},
       adsurl = {https://ui.adsabs.harvard.edu/abs/2022MNRAS.510.1128C}
}

@ARTICLE{RCM97,
       author = {{Ryu}, Dongsu and {Chakrabarti}, Sandip K. and {Molteni}, Diego},
        title = "{Zero-Energy Rotating Accretion Flows near a Black Hole}",
      journal = {\apj},
         year = 1997,
       volume = {474},
        pages = {378},
       adsurl = {https://ui.adsabs.harvard.edu/abs/1997ApJ...474..378R}
}

@ARTICLE{Markert79,
       author = {{Markert}, T.~H. and others},
        title = "{The MIT/OSO 7 catalog of X-ray sources: intensities, spectra, and
                  long-term variability}",
      journal = {\apjs},
         year = 1979,
       volume = {39},
        pages = {573},
       adsurl = {https://ui.adsabs.harvard.edu/abs/1979ApJS...39..573M}
}

@ARTICLE{Nath23,
       author = {{Nath}, Sujoy Kumar and {Debnath}, Dipak and {Chatterjee}, Kaushik and
                 {Jana}, Arghajit and {Chatterjee}, Debjit and {Bhowmick}, Riya},
        title = "{Accretion flow properties of MAXI J1910-057/Swift J1910.2-0546 during
                  its 2012-13 outburst}",
      journal = {Advances in Space Research},
         year = 2023,
       volume = {71},
        pages = {1045},
          doi = {10.1016/j.asr.2022.08.013},
archivePrefix = {arXiv},
       eprint = {2108.07456},
 primaryClass = {astro-ph.HE},
       adsurl = {https://ui.adsabs.harvard.edu/abs/2023AdSpR..71.1045N}
}

@ARTICLE{Bambi21,
       author = {{Bambi}, Cosimo and {Brenneman}, Laura W. and {Dauser}, Thomas and others},
        title = "{Towards Precision Measurements of Accreting Black Holes Using X-Ray
                  Reflection Spectroscopy}",
      journal = {Space Science Reviews},
         year = 2021,
        month = jul,
       volume = {217},
       number = {5},
          eid = {65},
        pages = {65},
          doi = {10.1007/s11214-021-00841-8},
archivePrefix = {arXiv},
       eprint = {2011.04792},
 primaryClass = {astro-ph.HE}
}

@ARTICLE{Brenneman06,
       author = {{Brenneman}, Laura W. and {Reynolds}, Christopher S.},
        title = "{Constraining Black Hole Spin via X-Ray Spectroscopy}",
      journal = {\apj},
         year = 2006,
       volume = {652},
        pages = {1028--1043}
}

@ARTICLE{Casella05,
       author = {{Casella}, Piergiorgio and {Belloni}, Tomaso and {Stella}, Luigi},
        title = "{The ABC of Low-Frequency Quasi-periodic Oscillations in Black Hole
                  Candidates: Analogies with Z Sources}",
      journal = {\apj},
         year = 2005,
        month = aug,
       volume = {629},
        pages = {403--407},
          doi = {10.1086/431174},
archivePrefix = {arXiv},
       eprint = {astro-ph/0504318},
 primaryClass = {astro-ph}
}

@ARTICLE{Galeev79,
       author = {{Galeev}, A.~A. and {Rosner}, R. and {Vaiana}, G.~S.},
        title = "{Structured coronae of accretion disks}",
      journal = {\apj},
         year = 1979,
        month = apr,
       volume = {229},
        pages = {318--326},
          doi = {10.1086/156957}
}

@ARTICLE{Haardt93,
       author = {{Haardt}, Francesco and {Maraschi}, Laura},
        title = "{X-Ray Spectra from Two-Phase Accretion Disks}",
      journal = {\apj},
         year = 1993,
       volume = {413},
        pages = {507--517}
}

@ARTICLE{Harrison13,
       author = {{Harrison}, Fiona A. and {Craig}, William W. and {Christensen}, Finn E. and others},
        title = "{The Nuclear Spectroscopic Telescope Array (NuSTAR) High-energy X-Ray Mission}",
      journal = {\apj},
         year = 2013,
        month = jun,
       volume = {770},
       number = {2},
          eid = {103},
        pages = {103},
          doi = {10.1088/0004-637X/770/2/103},
archivePrefix = {arXiv},
       eprint = {1301.7307},
 primaryClass = {astro-ph.IM}
}

@ARTICLE{Jana16,
       author = {{Jana}, Arghajit and {Debnath}, Dipak and {Chakrabarti}, Sandip K. and
                 {Mondal}, Santanu and {Molla}, Aslam Ali},
        title = "{Accretion Flow Dynamics of MAXI J1836-194 During Its 2011 Outburst
                  from TCAF Solution}",
      journal = {\apj},
         year = 2016,
        month = mar,
       volume = {819},
       number = {2},
          eid = {107},
        pages = {107},
          doi = {10.3847/0004-637X/819/2/107},
archivePrefix = {arXiv},
       eprint = {1601.02080},
 primaryClass = {astro-ph.HE}
}

@ARTICLE{Makishima86a,
       author = {{Makishima}, K. and {Maejima}, Y. and {Mitsuda}, K. and {Bradt}, H.~V. and
                 {Remillard}, R.~A. and {Tuohy}, I.~R. and {Hoshi}, R. and {Nakagawa}, M.},
        title = "{Simultaneous X-ray and optical observations of GX 339-4 in an X-ray
                  high state}",
      journal = {\apj},
         year = 1986,
        month = sep,
       volume = {308},
        pages = {635--643}
}

@ARTICLE{Mitsuda84,
       author = {{Mitsuda}, K. and {Inoue}, H. and {Koyama}, K. and {Makishima}, K. and
                 {Matsuoka}, M. and {Ogawara}, Y. and {Shibazaki}, N. and {Suzuki}, K. and
                 {Tanaka}, Y. and {Hirano}, T.},
        title = "{Energy Spectra of Low-Mass Binary X-Ray Sources Observed from Tenma}",
      journal = {\pasj},
         year = 1984,
       volume = {36},
       number = {4},
        pages = {741--759},
          doi = {10.1093/pasj/36.4.741}
}

@ARTICLE{MSC96,
       author = {{Molteni}, Diego and {Sponholz}, Hanno and {Chakrabarti}, Sandip K.},
        title = "{Resonance Oscillation of Radiative Shock Waves in Accretion Disks around
                  Compact Objects}",
      journal = {\apj},
         year = 1996,
        month = feb,
       volume = {457},
        pages = {805},
          doi = {10.1086/176775},
archivePrefix = {arXiv},
       eprint = {astro-ph/9508022},
 primaryClass = {astro-ph}
}

@ARTICLE{Reynolds14,
       author = {{Reynolds}, Christopher S.},
        title = "{Measuring Black Hole Spin Using X-Ray Reflection Spectroscopy}",
      journal = {Space Science Reviews},
         year = 2014,
        month = sep,
       volume = {183},
        pages = {277--294},
          doi = {10.1007/s11214-013-0006-6},
archivePrefix = {arXiv},
       eprint = {1302.3260},
 primaryClass = {astro-ph.HE}
}

@ARTICLE{Wardzinski00,
       author = {{Wardzi{\'n}ski}, Grzegorz and {Zdziarski}, Andrzej A.},
        title = "{Thermal synchrotron radiation and its Comptonization in compact X-ray sources}",
      journal = {\mnras},
         year = 2000,
       volume = {314},
        pages = {183},
          doi = {10.1046/j.1365-8711.2000.03297.x},
archivePrefix = {arXiv},
       eprint = {astro-ph/9911126},
 primaryClass = {astro-ph}
}

@ARTICLE{Zdziarski03,
       author = {{Zdziarski}, Andrzej A. and {Lubi{\'n}ski}, Piotr and {Gilfanov}, Marat and
                 {Revnivtsev}, Mike},
        title = "{Correlations between X-ray and radio spectral properties of accreting black holes}",
      journal = {\mnras},
         year = 2003,
       volume = {342},
       number = {2},
        pages = {355--372},
       adsurl = {https://ui.adsabs.harvard.edu/abs/2003MNRAS.342..355Z}
}

@ARTICLE{Zdziarski21,
       author = {{Zdziarski}, Andrzej A. and {Dzie{\l}ak}, Marta A. and {De Marco}, Barbara and
                 {Szanecki}, Micha{\l} and {Nied{\'z}wiecki}, Andrzej},
        title = "{Accretion Geometry in the Hard State of the Black Hole X-Ray Binary
                  MAXI J1820+070}",
      journal = {\apjl},
         year = 2021,
        month = mar,
       volume = {909},
       number = {1},
          eid = {L9},
        pages = {L9},
          doi = {10.3847/2041-8213/abe7ef},
archivePrefix = {arXiv},
       eprint = {2101.04482},
 primaryClass = {astro-ph.HE}
}

@INCOLLECTION{Cannizzo93,
       author = {{Cannizzo}, J.~K.},
        title = "{The limit cycle instability in dwarf nova accretion disks}",
    booktitle = {Accretion Disks in Compact Stellar Systems},
       editor = {{Wheeler}, J.~C.},
    publisher = {World Scientific},
         year = 1993,
        pages = {6},
        doi = {10.1142/9789814350976_0002}
}

@ARTICLE{Bhowmick25,
       author = {{Bhowmick}, Riya and {Nath}, Sujoy Kumar and {Debnath}, Dipak and {Chang}, Hsiang-Kuang},
        title = "{Jet X-Ray Properties of EXO 1846-031 During Its 2019 Outburst}",
      journal = {Universe},
         year = 2025,
        month = dec,
       volume = {11},
       number = {12},
          eid = {398},
        pages = {398},
          doi = {10.3390/universe11120398}
}

@MISC{Bhowmick26,
       author = {{Bhowmick}, R. and others},
         year = 2026,
         note = {in preparation}
}

@ARTICLE{Ebisawa96,
       author = {{Ebisawa}, Ken and {Titarchuk}, Lev and {Chakrabarti}, Sandip K.},
        title = "{On the Spectral Slopes of Hard X-Ray Emission from Black Hole Candidates}",
      journal = {\pasj},
         year = 1996,
        month = feb,
       volume = {48},
       number = {1},
        pages = {59--65},
          doi = {10.1093/pasj/48.1.59}
}

@ARTICLE{Francey71,
       author = {{Francey}, R.~J.},
        title = "{Position of Cen XR-2 (April 1967)}",
      journal = {Nature Physical Science},
         year = 1971,
       volume = {229},
        pages = {228}
}

@ARTICLE{Kimura19,
       author = {{Kimura}, Mariko and {Done}, Chris},
        title = "{Evolution of X-ray irradiation during the 1999-2000 outburst of the
                  black hole binary XTE J1859+226}",
      journal = {\mnras},
         year = 2019,
       volume = {482},
       number = {1},
        pages = {626--638},
          doi = {10.1093/mnras/sty2736}
}

@MISC{Makino87,
       author = {{Makino}, F.},
 howpublished = {IAU Circ.},
       number = {4342},
         year = 1987,
        pages = {1}
}

@InProceedings{Makishima86b,
author="Makishima, K.",
editor="Mason K. P. and Watson M. G. and White N. E.",
title=" Iron Lines from Galactic and Extragalactic X-ray Sources ",
booktitle="The Physics of Accretion onto Compact Objects",
year="1986",
publisher="Springer-Verlag",
address="Berlin Heidelberg New York",
volume = {266},
pages = {249}
}

@MISC{Negoro25,
       author = {{Negoro}, H. and {Fujiwara}, K. and {Nakajima}, M. and others},
        title = "{MAXI/GSC detection of renewed activity of the black hole X-ray binary GS 1354-64}",
 howpublished = {The Astronomer's Telegram},
       number = {17563},
         year = 2025,
        month = dec,
        pages = {1}
}

@ARTICLE{Russell19,
       author = {{Russell}, D.~M. and {Bramich}, D.~M. and {Lewis}, F. and others},
        title = "{Optical precursors to X-ray binary outbursts}",
      journal = {Astronomische Nachrichten},
         year = 2019,
       volume = {340},
        pages = {278},
archivePrefix = {arXiv},
       eprint = {1903.04519},
 primaryClass = {astro-ph.HE}
}

@MISC{Salvaggio26a,
       author = {{Salvaggio}, C. and {Grollimund}, N. and {Carotenuto}, F. and others},
        title = "{MeerKAT detects radio emission from black hole X-ray binary GS 1354-64}",
 howpublished = {The Astronomer's Telegram},
       number = {17582},
         year = 2026,
        month = jan,
        pages = {1}
}

@MISC{Salvaggio26b,
       author = {{Salvaggio}, C. and {Motta}, S.~E. and {Baglio}, M.~C. and others},
        title = "{The black hole X-ray binary GS 1354-64 in transition from the hard to the soft state}",
 howpublished = {The Astronomer's Telegram},
       number = {17638},
         year = 2026,
        month = jan,
        pages = {1}
}

@MISC{Salvaggio26c,
       author = {{Salvaggio}, C. and {Motta}, S.~E. and others},
         year = 2026,
         note = {in preparation}
}

@ARTICLE{Soleri13,
       author = {{Soleri}, P. and {Mu{\~n}oz-Darias}, T. and {Motta}, S. and {Belloni}, T. and
                 {Casella}, P. and {M{\'e}ndez}, M. and {Altamirano}, D. and {Linares}, M. and
                 {Wijnands}, R. and {Fender}, R. and {van der Klis}, M.},
        title = "{A complex state transition from the black hole candidate Swift J1753.5-0127}",
      journal = {\mnras},
         year = 2013,
        month = feb,
       volume = {429},
       number = {2},
        pages = {1244--1257}
}

@ARTICLE{Swinbanks87,
       author = {{Swinbanks}, D.},
       title = "{ Japan's Ginga (ASTRO-C) X-ray satellite }",
      journal = {Nature},
         year = 1987,
       volume = {326},
        pages = {322}
}

@ARTICLE{Vadawale01,
       author = {{Vadawale}, S.~V. and {Rao}, A.~R. and {Nandi}, A. and {Chakrabarti}, S.~K.},
        title = "{Observational evidence for mass ejection during soft X-ray dips in GRS 1915+105}",
      journal = {\aap},
         year = 2001,
       volume = {370},
        pages = {L17--L21},
          doi = {10.1051/0004-6361:20010318},
archivePrefix = {arXiv},
       eprint = {astro-ph/0103062},
 primaryClass = {astro-ph}
}

@ARTICLE{P21,
       author = {{Plotkin}, Richard M. and {Miller-Jones}, James C. A. and
                 {Bahramian}, A. and {Miller}, J. M. and
                 {Shahbaz}, T. and {Russell}, D. M.},
        title = "{Toward a Larger Sample of Radio Jets from Quiescent
                  Black Hole X-ray Binaries}",
      journal = {\mnras},
         year = 2021,
        month = apr,
       volume = {503},
       number = {3},
        pages = {3784--3795},
          doi = {10.1093/mnras/stab644},
archivePrefix = {arXiv},
       eprint = {2103.02178},
 primaryClass = {astro-ph.HE},
       adsurl = {https://ui.adsabs.harvard.edu/abs/2021MNRAS.503.3784P}
}

@ARTICLE{Gandhi19,
       author = {{Gandhi}, Poshak and {Rao}, Anjali and
                 {Johnson}, Michael A.~C. and {Paice}, John A. and
                 {Maccarone}, Thomas J.},
        title = "{Gaia Data Release 2 distances and peculiar velocities
                  for Galactic black hole transients}",
      journal = {\mnras},
         year = 2019,
        month = may,
       volume = {485},
       number = {2},
        pages = {2642--2655},
          doi = {10.1093/mnras/stz438},
archivePrefix = {arXiv},
       eprint = {1804.11349},
 primaryClass = {astro-ph.HE},
       adsurl = {https://ui.adsabs.harvard.edu/abs/2019MNRAS.485.2642G}
}

@ARTICLE{Sobczak99,
       author = {{Sobczak}, Gregory J. and {McClintock}, Jeffrey E. and
                 {Remillard}, Ronald A. and {Bailyn}, Charles D. and
                 {Orosz}, Jerome A.},
        title = "{RXTE Spectral Observations of the 1996--1997 Outburst
                  of the Microquasar GRO J1655-40}",
      journal = {\apj},
         year = 1999,
        volume = {520},
        number = {2},
        pages = {776--787},
          doi = {10.1086/307474},
archivePrefix = {arXiv},
       eprint = {astro-ph/9809195},
 primaryClass = {astro-ph},
       adsurl = {https://ui.adsabs.harvard.edu/abs/1999ApJ...520..776S}
}

@ARTICLE{Sobczak00,
       author = {{Sobczak}, Gregory J. and {McClintock}, Jeffrey E. and
                 {Remillard}, Ronald A. and {Cui}, Wei and
                 {Levine}, Alan M. and {Morgan}, Edward H. and
                 {Orosz}, Jerome A. and {Bailyn}, Charles D.},
        title = "{Complete RXTE Spectral Observations of the Black Hole
                  X-ray Nova XTE J1550-564}",
      journal = {\apj},
         year = 2000,
        month = dec,
       volume = {544},
       number = {2},
        pages = {993--1015},
          doi = {10.1086/317229},
archivePrefix = {arXiv},
       eprint = {astro-ph/0005599},
 primaryClass = {astro-ph},
       adsurl = {https://ui.adsabs.harvard.edu/abs/2000ApJ...544..993S}
}

@ARTICLE{McClintock09,
       author = {{McClintock}, Jeffrey E. and {Remillard}, Ronald A. and
                 {Rupen}, Michael P. and {Torres}, M.~A.~P. and
                 {Steeghs}, D. and {Levine}, Alan M. and
                 {Orosz}, Jerome A.},
        title = "{The 2003 Outburst of the X-ray Transient H1743-322:
                  Comparisons with the Black Hole Microquasar XTE J1550-564}",
      journal = {\apj},
         year = 2009,
        month = jun,
       volume = {698},
       number = {2},
        pages = {1398--1421},
          doi = {10.1088/0004-637X/698/2/1398},
archivePrefix = {arXiv},
       eprint = {0705.1034},
 primaryClass = {astro-ph},
       adsurl = {https://ui.adsabs.harvard.edu/abs/2009ApJ...698.1398M}
}

@ARTICLE{Stella98,
       author = {{Stella}, Luigi and {Vietri}, Mario},
        title = "{Lense-Thirring Precession and Quasi-periodic
                  Oscillations in Low-Mass X-Ray Binaries}",
      journal = {\apjl},
         year = 1998,
        volume = {492},
        number = {1},
        pages = {L59--L62},
          doi = {10.1086/311075},
       adsurl = {https://ui.adsabs.harvard.edu/abs/1998ApJ...492L..59S}
}

@ARTICLE{Ingram09,
       author = {{Ingram}, Adam and {Done}, Chris and
                 {Fragile}, P.~Chris},
        title = "{Low-frequency quasi-periodic oscillations spectra
                  and Lense-Thirring precession}",
      journal = {\mnras},
         year = 2009,
        month = jul,
       volume = {397},
       number = {1},
        pages = {L101--L105},
          doi = {10.1111/j.1745-3933.2009.00693.x},
       adsurl = {https://ui.adsabs.harvard.edu/abs/2009MNRAS.397L.101I}
}

@ARTICLE{Ingram16,
       author = {{Ingram}, Adam and {van der Klis}, Michiel and {Middleton}, Matthew and
                 {Done}, Chris and {Altamirano}, Diego and {Heil}, Lucy and
                 {Uttley}, Phil and {Axelsson}, Magnus},
        title = "{A quasi-periodic modulation of the iron line centroid energy in the
                  black hole binary H1743-322}",
      journal = {\mnras},
         year = 2016,
        month = sep,
       volume = {461},
       number = {2},
        pages = {1967--1980},
          doi = {10.1093/mnras/stw1245},
archivePrefix = {arXiv},
       eprint = {1607.02866},
 primaryClass = {astro-ph.HE},
       adsurl = {https://ui.adsabs.harvard.edu/abs/2016MNRAS.461.1967I}
}

@article{Brocksopp2002,
  author  = {Brocksopp, C. and Fender, R. P. and McCollough, M.
             and Pooley, G. G. and Rupen, M. P. and Hjellming, R. M.
             and de la Force, C. J. and Spencer, R. E.
             and Muxlow, T. W. B. and Garrington, S. T.
             and Trushkin, S.},
  title   = {Initial low/hard state, multiple jet ejections and
             X-ray/radio correlations during the outburst of
             {XTE J1859+226}},
  journal = {MNRAS},
  year    = {2002},
  volume  = {331},
  pages   = {765--775},
  adsurl  = {https://ui.adsabs.harvard.edu/abs/2002MNRAS.331..765B}
}
\bibliographystyle{aasjournalv7.1}

\begin{table*}
\addtolength{\tabcolsep}{-3.5pt} 
\vskip -0.3cm
\centering
\caption{Model Fitted Temporal and Spectral Parameters}
\label{table_master}
\scriptsize
\addtolength{\tabcolsep}{-1pt}
\renewcommand{\arraystretch}{0.9}  
\begin{tabular}{lcc|cccc|ccccccccccccc}  
 \hline
\multicolumn{3}{c}{\textsc{Data Info.}} & \multicolumn{4}{c}{\textsc{QPO Fitted Result}} &  \multicolumn{11}{c}{\textsc{Spectral Fitted Result}}  \\
 \hline
Sr.& ObsID &  MJD & $\nu_{QPO}$ & Q & rms(\%) &  type & $T_{\rm in}$ & $\Gamma_{\rm th}$ & $kT_{\rm e}$ & $f_{\rm cov}$ & $LinE$ & $q*$ &  $\Gamma_{\rm PL}$ & $\chi^2/DOF$ & $F_{\rm bol}$ & $F_{nth}$ & $F_{dbb}$ & $F_{RL}$ & $F_{PL}$\\
     &     &     &(Hz) &     &     &     &(keV)&     &(keV) &     &(keV)& &  &  & \multicolumn{5}{c}{\textsc{($10^{-9}~{\rm erg}~{\rm cm}^{-2}~{\rm s}^{-1}$})} \\
(1) & (2) & (3) & (4) & (5) & (6) & (7) & (8) & (9) & (10) & (11) & (12) & (13)& (14)& (15)& (16)& (17)& (18)& (19) & (20)\\
\hline
 1 &  NX2002 &  61053.60 &  $0.07^{\pm0.002}$ &  $13.7$ &  $3.21$ &  C &  $0.25^f$ &  $1.69^{\pm0.01}$ &  $8.0^{\pm0.06}$ &  $0.89^{\pm0.01}$ &  $6.18^{\pm0.03}$ &  $1.89^{\pm0.07}$ &  $1.34^{\pm0.01}$ &  $994/901$ &  $11.4$ &  $7.78$ &  $0.21$ &  $0.06$ &  $3.36$ \\
2  & HX0101 & 61058.17 & $0.26^{\pm0.01}$ & $14.9$ & $8.13$ & C &  --& --&-- & -- & --& --&-- & -- & --& --&--& --&--\\ 
3  & HX0102 & 61058.32 & $0.28^{\pm0.01}$ & $18.3$ & $3.56$ & C & $0.58^{\pm0.06}$ & $1.71^{\pm0.01}$ & $8.9^{\pm0.78}$ & $0.92^{\pm0.01}$ & --& --&--                                                         & $838/887$ & $9.44$ &$8.78$ & $0.66$ & --&--\\
4  & HX0103 & 61058.44 & $0.28^{\pm0.02}$ & $14.9$ & $4.40$ & C & $0.62^{\pm0.03}$ & $1.72^{\pm0.02}$ & $9.4^{\pm0.85}$ & $0.93^{\pm0.01}$ & --& --&--                                                         & $761/784$ & $9.94$ &$9.02$ & $0.92$ & --&--\\
 5 &  NX2004 &  61061.54 &  $1.46^{\pm0.01}$ &  $15.8$ &  $3.93$ &  C &  $0.41^{\pm0.03}$ &  $2.00^{\pm0.01}$ &  $7.8^{\pm0.11}$ &  $0.97^{\pm0.02}$ &  $6.19^{\pm0.07}$ &  $2.01^{\pm0.06}$ &  $1.58^{\pm0.01}$ &  $846/752$ &  $11.7$ &  $7.92$ &  $1.29$ &  $0.07$ &  $2.47$ \\
 6 &  NX2006 &  61064.69 &  $1.73^{\pm0.01}$ &  $15.6$ &  $5.28$ &  C &  $0.49^{\pm0.04}$ &  $2.06^{\pm0.01}$ &  $8.6^{\pm0.15}$ &  $0.92^{\pm0.01}$ &  $6.24^{\pm0.09}$ &  $2.06^{\pm0.11}$ &  $1.16^{\pm0.06}$ &  $779/705$ &  $11.5$ &  $7.94$ &  $1.93$ &  $0.06$ &  $1.58$ \\
7  & HY0101 & 61066.08 & $2.33^{\pm0.05}$ & $20.0$ & $7.68$& C & -- & --& --&-- & -- & --&-- & -- & --& --&--& --& --\\
8  & HY0102 & 61066.22 & $2.59^{\pm0.03}$ & $21.1$ & $7.33$ & C & $0.50^{\pm0.02}$ & $2.12^{\pm0.03}$ & $14^{\pm3.3}$    & $0.47^{\pm0.02}$ & ---& --&--                                                       & $699/787$ & $12.1$ &$8.01$ & $4.09$ & --&-- \\
9  & HY0103 & 61066.37 & $2.72^{\pm0.03}$ & $19.4$ & $11.0$ & C & $0.54^{\pm0.03}$ & $2.12^{\pm0.02}$ & $20^{f}$       & $0.51^{\pm0.03}$  & -- & --&--                                                        & $748/840$ & $11.6$ &$7.41$ & $4.19$ & --&--\\
10 & HY0104 & 61066.52 & $2.77^{\pm0.04}$ & $18.6$ & $10.5$ & C & $0.59^{\pm0.03}$ & $2.10^{\pm0.02}$ & $15^{f}$       & $0.47^{\pm0.02}$  & -- & --&--                                                        & $773/808$ & $11.6$ &$7.19$ & $4.41$ & --&--\\
11 & HX0201 & 61067.14 & $3.36^{\pm0.09}$ & $21.4$ & $4.87$ & C & $0.61^{\pm0.04}$ & $2.23^{\pm0.03}$ & $20^{f}$       & $0.54^{\pm0.03}$  & -- & --&--                                                        & $766/768$ & $13.0$ &$7.65$ & $5.35$ & --&--\\
12 & HX0202 & 61067.28 & $3.53^{\pm0.11}$ & $23.4$ & $3.68$ & C & $0.62^{\pm0.03}$ & $2.23^{\pm0.02}$ & $20^{f}$       & $0.54^{\pm0.03}$  & -- & --&--                                                        & $811/806$ & $12.7$ &$7.46$ & $5.24$ & --&-- \\
13 & HX0203 & 61067.41 & $3.92^{\pm0.08}$ & $19.7$ & $12.8$ & C & -- & --& --&-- & -- & --& --&-- & -- & --& --&-- \\

14 & HY0301 & 61068.33 & $5.76^{\pm0.08}$ & $31.2$ & $5.02$ & B &                        $0.63^{\pm0.03}$ & $2.25^{\pm0.02}$ &  $25^{f}$      & $0.57^{\pm0.03}$   & --& --&--                                                      & $780/793$ & $12.6$ &$7.35$ & $5.25$ & -- & -- \\
15 & HY0302 & 61068.49 & -- & -- & -- & -- &                        $0.79^{\pm0.04}$ & $2.42^{\pm0.03}$ &  $20^{f}$      & $0.56^{\pm0.05}$   & --& --&--                                                      & $760/766$ & $16.7$ &$7.62$ & $9.08$ & -- & -- \\
16 & HY0303 & 61068.61 & -- & -- & -- & -- &                        $0.92^{\pm0.04}$ & $2.43^{\pm0.04}$ &  $30^{f}$      & $0.48^{\pm0.06}$   & --& --&--                                                      & $827/801$ & $19.2$ &$7.40$ & $11.8$ & --&--\\
17 & HY0601 & 61069.32 & -- & -- & -- & -- &                        $0.98^{\pm0.02}$ & $2.51^{\pm0.04}$ &  $30^{f}$      & $0.50^{\pm0.04}$   & --& --&--                                                      & $944/916$ & $26.0$ &$9.20$ & $16.8$ & --&--\\
18 & HY0602 & 61069.47 & -- & -- & -- & -- &                        $0.98^{\pm0.03}$ & $2.66^{\pm0.04}$ &  $30^{f}$      & $0.59^{\pm0.06}$   & --& --&--                                                      & $846/881$ & $26.0$ &$9.20$ & $16.8$ & --&--\\
19 & HY0603 & 61069.60 & -- & -- & -- & -- &                        $0.98^{\pm0.03}$ & $2.66^{\pm0.04}$ &  $30^{f}$      & $0.59^{\pm0.06}$   & --& --&--                                                      & $846/881$ & $30.8$ &$14.0$ & $16.8$ & --&--\\
 20 &  $NY002^{a}$ &  61070.25 &   -- &  -- &  -- &  -- &                  $0.95^{\pm0.03}$ &  $2.54^{\pm0.02}$ &   $54^{\pm13.2}$ &  $0.78^{\pm0.07}$  &  $6.29^{\pm0.09}$ &  $2.00^{\pm0.10}$ &   --                         &  $566/561$ &  $19.3$ & $8.26$ &  $10.9$ &  $0.07$ &  -- \\   
 21 &  $NY002^{b}$ &  61070.48 &  $6.14^{\pm0.18}$ &  $14.3$ &  $12.4$ &  C &  $0.93^{\pm0.03}$ &  $2.50^{\pm0.04}$ &   $41^{\pm13.1}$  &  $0.74^{\pm0.09}$  &  $6.42^{\pm0.16}$ &  $1.95^{\pm0.13}$ &  --                     &  $515/474$ &  $18.1$ & $8.05$ &  $10.0$ &  $0.08$ &  -- \\
22 & HY0701 & 61070.52 & $6.16^{\pm0.29}$ & $14.3$ & $5.35$ & C & $0.69^{\pm0.03}$ & $2.43^{\pm0.02}$ & $40^{f}$       & $0.75^{\pm0.04}$  & -- & --&--                                                        & $816/852$ & $18.7$ &$10.0$ & $8.67$  & --&--\\
23 & HY0702 & 61070.67 & $5.07^{\pm0.31}$ & $17.3$ & $4.00$ & C & $0.67^{\pm0.03}$ & $2.42^{\pm0.02}$ & $40^{f}$       & $0.73^{\pm0.04}$  & -- & --&--                                                        & $892/884$ & $16.5$ &$8.90$ & $7.60$  & --&--\\
24 & HY0703 & 61070.79 & -- & -- & -- & -- &                      $0.70^{\pm0.03}$ & $2.45^{\pm0.02}$ & $40^{f}$       & $0.72^{\pm0.04}$  & -- & --&--                                                        & $906/885$ & $16.1$ &$8.30$ & $7.80$  & --&--\\
25 & HY0801 & 61071.51 & $3.63^{\pm0.05}$ & $20.2$ & $8.32$ & C & $0.56^{\pm0.03}$ & $2.26^{\pm0.03}$ & $40^{f}$       & $0.65^{\pm0.03}$  & -- & --&--                                                        & $815/867$ & $13.8$ &$8.55$ & $5.25$  & --&--\\
26 & HY0802 & 61071.66 & $3.50^{\pm0.04}$ & $17.8$ & $11.4$ & C & $0.56^{\pm0.03}$ & $2.28^{\pm0.03}$ & $40^{f}$       & $0.66^{\pm0.03}$  & -- & --&--                                                        & $832/874$ & $13.1$ &$8.11$ & $4.99$  & --&--\\
27 & HY0803 & 61071.78 & $3.26^{\pm0.11}$ & $19.4$ & $5.54$ & C & $0.57^{\pm0.04}$ & $2.18^{\pm0.04}$ & $40^{f}$       & $0.57^{\pm0.04}$  & -- & --&--                                                        & $637/741$ & $13.0$ &$8.20$ & $4.80$ & -- & --\\
28 & HY0901 & 61072.70 & $3.64^{\pm0.06}$ & $21.1$ & $6.66$ & C & $0.57^{\pm0.02}$ & $2.29^{\pm0.02}$ & $40^{f}$       & $0.69^{\pm0.03}$  & -- & --&--                                                        & $853/924$ & $12.8$ &$7.93$ & $4.87$  & --&--\\
29 & HY0902 & 61072.84 & $3.79^{\pm0.07}$ & $20.2$ & $7.10$ & C & $0.55^{\pm0.03}$ & $2.27^{\pm0.03}$ & $40^{f}$       & $0.66^{\pm0.03}$  & -- & --&--                                                        & $897/882$ & $13.4$ &$8.34$ & $5.06$  & --&--\\
30 & HY0903 & 61072.99 & $3.61^{\pm0.06}$ & $21.1$ & $7.19$& C & -- & --& --&-- & -- & --& --&-- & -- & --&--& --& --\\
31 & HY1001 & 61073.75 & $3.16^{\pm0.06}$ & $20.2$ & $6.13$ & C & $0.54^{\pm0.02}$ & $2.19^{\pm0.02}$ & $30^{f}$       & $0.58^{\pm0.02}$  & -- & --&--                                                        & $816/906$ & $12.1$ &$7.86$ & $4.24$ & -- & --\\
32 & HY1002 & 61073.90 & $3.32^{\pm0.06}$ & $23.1$ & $5.74$ & C & $0.53^{\pm0.02}$ & $2.19^{\pm0.03}$ & $20^{f}$       & $0.52^{\pm0.02}$  & -- & --&--                                                        & $844/878$ & $12.1$ &$7.76$ & $4.44$ & -- & --\\
33 & HY1101 & 61074.94 & $3.42^{\pm0.04}$ & $17.8$ & $9.66$ & C & $0.54^{\pm0.02}$ & $2.21^{\pm0.02}$ & $20^{f}$       & $0.54^{\pm0.02}$  & -- & --&--                                                        & $835/907$ & $12.0$ &$7.65$ & $4.35$ & -- & --\\
34 & HY1102 & 61075.08 & $3.61^{\pm0.07}$ & $16.7$ & $9.34$ & C & $0.55^{\pm0.03}$ & $2.24^{\pm0.03}$ & $20^{f}$       & $0.54^{\pm0.03}$  & -- & --&--                                                        & $826/878$ & $12.1$ &$7.43$ & $4.67$ & -- & --\\
35 & HY1103 & 61075.21 & $4.00^{\pm0.09}$ & $20.8$ & $8.98$ & C & $0.59^{\pm0.04}$ & $2.24^{\pm0.04}$ & $20^{f}$       & $0.52^{\pm0.03}$  & -- & --&--                                                        & $827/768$ & $12.4$ &$7.26$ & $5.14$ & -- & --\\
36 & HY1301 & 61077.10 & $4.59^{\pm0.12}$ & $16.9$ & $8.43$ & C & $0.65^{\pm0.03}$ & $2.27^{\pm0.03}$ & $20^{f}$       & $0.52^{\pm0.03}$  & -- & --&--                                                        & $804/785$ & $12.4$ &$6.81$ & $5.59$ & -- & --\\
37 & HY1302 & 61077.23 & $4.46^{\pm0.54}$ & $20.5$ & $3.10$ & C & $0.67^{\pm0.03}$ & $2.28^{\pm0.04}$ & $20^{f}$       & $0.53^{\pm0.04}$  & -- & --&--                                                        & $764/809$ & $13.5$ &$7.28$ & $6.22$ & -- & --\\
38 & HY1401 & 61078.30 & $4.25^{\pm0.21}$ & $19.4$ & $6.01$ & C & $0.63^{\pm0.03}$ & $2.28^{\pm0.04}$ & $30^{f}$       & $0.58^{\pm0.04}$  & -- & --&--                                                        & $812/790$ & $13.3$ &$7.54$ & $5.76$ & -- & --\\
39 & HY1402 & 61078.44 & $4.33^{\pm0.35}$ & $24.3$ & $2.78$ & C & $0.60^{\pm0.03}$ & $2.24^{\pm0.04}$ & $20^{f}$       & $0.53^{\pm0.03}$  & -- & --&--                                                        & $829/888$ & $12.5$ &$7.29$ & $5.21$ & -- & --\\
40 & HY1403 & 61078.56 & $3.86^{\pm0.11}$ & $19.4$ & $9.23$ & C & $0.60^{\pm0.03}$ & $2.18^{\pm0.05}$ & $20^{f}$       & $0.47^{\pm0.03}$  & -- & --&--                                                        & $779/766$ & $11.6$ &$6.77$ & $4.83$ & -- & --\\
41 & HY1901 & 61079.07 & $4.75^{\pm0.20}$ & $19.8$ & $5.07$ & C & $0.63^{\pm0.02}$ & $2.31^{\pm0.03}$ & $20^{f}$       & $0.56^{\pm0.03}$  & -- & --&--                                                        & $864/871$ & $12.7$ &$7.05$ & $5.65$ & -- & --\\
42 & HY1903 & 61079.34 & $5.02^{\pm0.10}$ & $7.07$ & $8.51$ & C & $0.69^{\pm0.02}$ & $2.38^{\pm0.03}$ & $20^{f}$       & $0.58^{\pm0.03}$  & -- & --&--                                                        & $959/901$ & $12.8$ &$6.51$ & $6.29$ & -- & --\\
43 & HY2001 & 61080.32 & $4.95^{\pm0.11}$ & $6.55$ & $7.80$ & C & $0.66^{\pm0.02}$ & $2.30^{\pm0.03}$ & $20^{f}$       & $0.54^{\pm0.03}$  & -- & --&--                                                        & $852/907$ & $12.8$ &$6.87$ & $5.93$ & -- & --\\
44 & HY2002 & 61080.48 & $5.46^{\pm0.75}$ & $19.9$ & $12.1$& C  & $0.73^{\pm0.02}$ & $2.34^{\pm0.03}$ & $20^{f}$       & $0.48^{\pm0.03}$  & -- & --&--                                                        & $1027/901$& $12.7$ &$5.96$ & $6.74$ & -- & --\\
45 & HY2003 & 61080.61 & $5.12^{\pm0.11}$ & $6.89$ & $9.97$& C  & $0.66^{\pm0.03}$ & $2.36^{\pm0.05}$ & $20^{f}$       & $0.57^{\pm0.04}$  & -- & --&--                                                        & $808/855$ & $12.6$ &$6.65$ & $5.95$ & -- & --\\
46 & HY1501 & 61081.18 & $4.87^{\pm0.09}$ & $6.54$ & $9.86$& C &  -- & --& --&-- & -- & --& --&-- & --& --&--& --& --\\
47 & HY1502 & 61081.33 & $4.91^{\pm0.10}$ & $6.58$ & $7.66$&  C  & $0.66^{\pm0.02}$ & $2.35^{\pm0.03}$ & $30^{f}$      & $0.61^{\pm0.03}$  & --& --&--                                                          & $941/924$ & $12.6$ &$6.77$ & $5.83$ & -- & --\\
48 & HY1503 & 61081.46 & $5.48^{\pm0.35}$ & $12.8$ & $6.44$ & C & $0.69^{\pm0.03}$ & $2.33^{\pm0.04}$ & $30^{f}$       & $0.57^{\pm0.04}$  & --& --&--                                                          & $822/872$ & $12.8$ &$6.67$ & $6.13$ & -- & --\\
 49 &  NZ1002 &  61082.76 &  $4.67^{\pm0.04}$ &  $7.67$ &  $8.22$ &  C &  $0.87^{\pm0.02}$ &  $2.37^{\pm0.03}$ &  $32^{\pm7.5}$       &  $0.67^{\pm0.06}$  &  $6.35^{\pm0.08}$ &   $1.96^{\pm0.13}$ &   --                         &  $525/523$ &  $11.7$ & $5.77$ &  $5.90$ &  $0.05$ &  -- \\
50 & HY2101 & 61083.09 & $4.77^{\pm0.18}$ & $8.42$ & $8.14$ & C & $0.65^{\pm0.02}$ & $2.31^{\pm0.03}$ & $20^{f}$       & $0.57^{\pm0.03}$  & --& --&--                                                          & $893/891$ & $12.0$ &$6.52$ & $5.48$ & -- & --\\
51 & HY2102 & 61083.24 & $5.14^{\pm0.12}$ & $9.99$ & $8.89$ & C & $0.67^{\pm0.02}$ & $2.35^{\pm0.03}$ & $30^{f}$       & $0.58^{\pm0.03}$  & --& --&--                                                          & $921/907$ & $12.1$ &$6.37$ & $5.73$ & -- & --\\
52 & HY2103 & 61083.41 & $5.03^{\pm0.17}$ & $11.7$ & $5.49$ & C & $0.64^{\pm0.02}$ & $2.35^{\pm0.03}$ & $40^{f}$       & $0.64^{\pm0.03}$  & --& --&--                                                          & $914/907$ & $12.0$ &$6.55$ & $5.45$ & -- & --\\
53 & HY2201 & 61085.11 & $6.25^{\pm0.30}$ & $13.8$ & $3.56$ & C & $0.71^{\pm0.02}$ & $2.44^{\pm0.03}$ & $40^{f}$       & $0.63^{\pm0.03}$  & --& --&--                                                          & $857/892$ & $13.1$ &$6.32$ & $6.78$ & -- & --\\
 54 &  NZ1004 &  61085.17 &  $5.81^{\pm0.13}$ &  $4.51$ &  $12.6$ &  B &  $0.94^{\pm0.02}$ &  $2.42^{\pm0.01}$ &  $50^{f}$       &  $0.63^{\pm0.02}$  & $6.63^{\pm0.14}$ &  $1.97^{\pm0.18}$ &  --                            &  $492/485$ &  $12.5$ & $5.43$ &  $7.07$ &  $0.05$ &  -- \\
55 & HY2301 & 61086.88 & -- & -- & -- & -- &                        $0.90^{\pm0.01}$ & $2.58^{\pm0.02}$ & $30^{f}$       & $0.45^{\pm0.02}$  & --& --&--                                                        & $1074/907$& $17.6$ &$5.70$ & $11.9$ & -- & --\\
56 & HY2302 & 61087.08 & -- & -- & -- & -- &                        $0.89^{\pm0.01}$ & $2.59^{\pm0.03}$ & $25^{f}$       & $0.43^{\pm0.03}$  & --& --&--                                                        & $981/907$ & $17.3$ &$5.40$ & $11.9$ & -- & --\\
57 & HY2502 & 61088.92 & -- & -- & -- & -- &                        $0.75^{\pm0.01}$ & $2.44^{\pm0.02}$ & $30^{f}$       & $0.52^{\pm0.02}$  & --& --&--                                                        & $883/907$ & $12.3$ &$5.33$ & $6.97$ & -- & --\\
58 & HY2601 & 61090.49 & $4.81^{\pm0.20}$ & $10.8$ & $4.84$ & B & $0.66^{\pm0.02}$ & $2.31^{\pm0.03}$ & $25^{f}$       & $0.49^{\pm0.02}$  & --& --&--                                                          & $929/907$ & $9.61$ &$4.89$ & $4.72$ & -- & --\\
59 & HY2602 & 61090.64 & $5.05^{\pm0.20}$ & $5.63$ & $6.76$ & B & $0.65^{\pm0.02}$ & $2.33^{\pm0.03}$ & $25^{f}$       & $0.52^{\pm0.03}$  & --& --&--                                                          & $821/906$ & $9.84$ &$5.10$ & $4.74$ & -- & --\\
60 & HY2603 & 61090.76 & -- & -- & -- & -- &                        $0.68^{\pm0.02}$ & $2.36^{\pm0.04}$ & $25^{f}$       & $0.48^{\pm0.03}$  & --& --&--                                                        & $794/836$ & $10.4$ &$4.94$ & $5.46$ & -- & --\\
\hline

\end{tabular}
\noindent{
\leftline{NX=91201302, NY=91202308, NZ=91202311, HX=P081404700, HY=P081404800 are pre-fixes of the NuSTAR and HXMT observation IDs.}
\leftline{$\nu_{QPO}$, Q ($=FWHM/\nu_{QPO}$), \% of rms define nature of the QPOs. DOF means degree of freedom.}
\leftline{For spectral fitting \texttt{Tbabs} N$_{\rm H} = 7\times 10^{21}~cm^{-2}$, \texttt{relline} {$R_{\rm in} =  -1 r_g (=GM/c^2)$}, $R_{\rm out} = 400 r_g$,  $a_k=0.998$, and $i=66^\circ$ are kept frozen.}
\leftline{Here, T$_{\rm in}$ means \texttt{diskbb} disk temparature; $\Gamma_{\rm th}$, kT$_e$, and f$_{\rm cov}$ mean \texttt{thcomp} photon index, electron temperature, and covering fraction respectively;} 
\leftline{$LinE$, and $q*$ mean line energy and emissivity index respectively; $\Gamma_{PL}$ means photon index of additional powerlaw, required to fit excess flux above $50$~keV.}
\leftline{F$_{\rm bol}$, F$_{\rm nth}$, F$_{\rm dbb}$, F$_{\rm RL}$, F$_{\rm PL}$ mean $1-100$~keV bolometric, and thcomp, diskbb, relline, additional powerlaw respectively model component fluxes.}
\leftline{$^f$ Unable to constrain errors, \texttt{diskbb} T$_{\rm in}$ for first observation and \texttt{thcomp} kT$_e$ for a few observations are frozen at their best fitted values.}
\leftline{Note, due to lack of good s/n data above $>7$~keV spectral fitting was not performed for few \textit{Insight}-HXMT observations.}
\leftline{$^a$ initial part ($<25$~ks) and $^b$ remianing part of the same \textit{NuSTAR} observation.}
}
\end{table*}

\addtocounter{table}{-1}
\begin{table*}
\addtolength{\tabcolsep}{-3.0pt} 
\vskip -0.3cm
\centering
\caption{(continued): Model Fitted Temporal and Spectral Parameters}
\label{table_master_cont}
\scriptsize
\addtolength{\tabcolsep}{-1pt}
\renewcommand{\arraystretch}{0.9}  
\begin{tabular}{lcc|cccc|ccccccccccccc}
 \hline
\multicolumn{3}{c}{\textsc{Data Info.}} & \multicolumn{4}{c}{\textsc{QPO Fitted Result}} &  \multicolumn{11}{c}{\textsc{Spectral Fitted Result}}  \\
 \hline
Sr.& ObsID &  MJD & $\nu_{QPO}$ & Q & rms(\%) &  type & $T_{\rm in}$ & $\Gamma_{\rm th}$ & $kT_{\rm e}$ & $f_{\rm cov}$ & $LinE$ & $q*$ &  $\Gamma_{\rm PL}$ & $\chi^2/DOF$ & $F_{\rm bol}$ & $F_{nth}$ & $F_{dbb}$ &  $F_{RL}$ & $F_{PL}$\\
     &     &     &(Hz) &     &     &     &(keV)&     &(keV) &     &(keV)& &  &  & \multicolumn{5}{c}{\textsc{($10^{-9}~{\rm erg}~{\rm cm}^{-2}~{\rm s}^{-1}$})} \\
(1) & (2) & (3) & (4) & (5) & (6) & (7) & (8) & (9) & (10) & (11) & (12) & (13)& (14)& (15)& (16)& (17)& (18)& (19) & (20)\\
\hline
 61 &  NZ1006 &  61091.13 &  -- &  -- &  -- &  -- &                         $1.00^{\pm0.01}$ &  $2.38^{\pm0.02}$ &  $45^{f}$       &  $0.43^{\pm0.01}$  &  $6.64^{\pm0.13}$ &  $2.07^{\pm0.15}$ &  --                         &  $527/472$ &  $12.9$ & $4.45$ &  $8.38$ &  $0.06$ &  -- \\
62 & HY2701 & 61092.30 & -- & -- & -- & -- &                        $0.88^{\pm0.01}$ & $2.49^{\pm0.03}$ & $30^{f}$       & $0.37^{\pm0.02}$  & --& --&--                                                        & $1018/890$& $14.6$ &$4.50$ & $10.1$ & -- & --\\
63 & HY2702 & 61092.50 & -- & -- & -- & -- &                        $0.86^{\pm0.01}$ & $2.55^{\pm0.04}$ & $30^{f}$       & $0.44^{\pm0.03}$  & --& --&--                                                        & $875/907$ & $14.3$ &$4.80$ & $9.50$ & -- & --\\
 64 &  NZ1008 &  61093.65 &  $5.63^{\pm0.09}$ &  $2.68$ &  $27.2$ &  B &        $1.04^{\pm0.01}$ &  $2.44^{\pm0.02}$ &  $45^{f}$      &  $0.40^{\pm0.01}$  &  $6.69^{\pm0.16}$ &  $2.01^{\pm0.15}$ &  --                      &  $539/482$ &  $13.8$ & $4.27$ &  $9.49$ &  $0.05$ &  -- \\
65 & HY2801 & 61095.02 & -- & -- & -- & -- &        $0.79^{\pm0.02}$ & $2.43^{\pm0.04}$ & $50^{f}$       & $0.45^{\pm0.04}$  & --& --&--                                                          & $775/788$ & $10.7$ &$4.00$ & $6.70$ & --& --\\
66 & HY2802 & 61095.16 & -- & -- & -- & -- &        $0.74^{\pm0.02}$ & $2.41^{\pm0.03}$ & $40^{f}$       & $0.48^{\pm0.03}$  & --& --&--                                                          & $902/849$ & $10.1$ &$4.21$ & $5.89$ & --& --\\
67 & HY2803 & 61095.29 & -- & -- & -- & -- &        $0.82^{\pm0.02}$ & $2.39^{\pm0.03}$ & $40^{f}$       & $0.39^{\pm0.03}$  & --& --&--                                                          & $865/880$ & $10.1$ &$4.21$ & $5.89$ & --& --\\
68 & HY2901 & 61096.34 & -- & -- & -- & -- &        $0.83^{\pm0.03}$ & $2.58^{\pm0.04}$ & $30^{f}$       & $0.49^{\pm0.05}$  & --& --&--                                                          & $708/747$ & $12.2$ &$4.30$ & $7.90$ & --& --\\
69 & HY2903 & 61096.61 & -- & -- & -- & -- &        $0.84^{\pm0.02}$ & $2.55^{\pm0.05}$ & $30^{f}$       & $0.34^{\pm0.04}$  & --& --&--                                                          & $789/787$ & $12.5$ &$3.49$ & $9.01$ & --& --\\
70 & HY3001 & 61097.92 & -- & -- & -- & -- &        $0.76^{\pm0.03}$ & $2.39^{\pm0.11}$ & $50^{f}$       & $0.45^{\pm0.06}$  & --& --&--                                                          & $583/596$ & $10.4$ &$4.13$ & $6.27$ & --& --\\
71 & HY3002 & 61098.05 & -- & -- & -- & -- &        $0.79^{\pm0.01}$ & $2.40^{\pm0.04}$ & $30^{f}$       & $0.41^{\pm0.03}$  & --& --&--                                                          & $844/845$ & $11.3$ &$4.32$ & $6.98$ & --& --\\
72 & HY3003 & 61098.19 & -- & -- & -- & -- &        $0.79^{\pm0.02}$ & $2.36^{\pm0.04}$ & $30^{f}$       & $0.38^{\pm0.03}$  & --& --&--                                                          & $762/799$ & $10.8$ &$4.11$ & $6.69$ & --& --\\
 73 &  NZ1010 &  61098.42 &  -- &  -- &  -- &  -- &         $0.99^{\pm0.01}$ &  $2.33^{\pm0.03}$ &  $35^{f}$       &  $0.37^{\pm0.01}$  & $6.72^{\pm0.16}$ &  $2.06^{\pm0.11}$ & \--                            &  $493/467$ &  $10.5$ & $3.57$ &  $6.87$ &  $0.05$ &  -- \\
74 & HY3101 & 61098.91 & -- & -- & -- & -- &        $0.71^{\pm0.02}$ & $2.35^{\pm0.04}$ & $40^{f}$       & $0.45^{\pm0.03}$  & --& --&--                                                          & $758/770$ & $8.81$ &$3.81$ & $5.00$ & --& --\\
75 & HY3102 & 61099.04 & -- & -- & -- & -- &        $0.71^{\pm0.02}$ & $2.32^{\pm0.04}$ & $40^{f}$       & $0.45^{\pm0.03}$  & --& --&--                                                          & $785/764$ & $8.78$ &$3.93$ & $4.85$ & --& --\\
76 & HY3103 & 61099.18 & -- & -- & -- & -- &        $0.67^{\pm0.02}$ & $2.39^{\pm0.04}$ & $40^{f}$       & $0.52^{\pm0.03}$  & --& --&--                                                          & $808/790$ & $8.68$ &$4.18$ & $4.50$ & --& --\\
77 & HY3201 & 61099.90 & -- & -- & -- & -- &        $0.82^{\pm0.02}$ & $2.32^{\pm0.05}$ & $40^{f}$       & $0.33^{\pm0.03}$  & --& --&--                                                          & $671/707$ & $10.2$ &$3.60$ & $6.60$ & --& --\\
78 & HY3202 & 61100.02 & -- & -- & -- & -- &        $0.79^{\pm0.01}$ & $2.41^{\pm0.04}$ & $40^{f}$       & $0.42^{\pm0.02}$  & --& --&--                                                          & $867/852$ & $10.3$ &$3.87$ & $6.43$ & --& --\\
79 & HY3203 & 61100.16 & -- & -- & -- & -- &        $0.82^{\pm0.02}$ & $2.46^{\pm0.04}$ & $40^{f}$       & $0.44^{\pm0.03}$  & --& --&--                                                          & $802/824$ & $10.9$ &$3.88$ & $7.02$ & --& --\\
80 & HY3301 & 61100.71 & -- & -- & -- & -- &        $0.81^{\pm0.02}$ & $2.48^{\pm0.04}$ & $40^{f}$       & $0.44^{\pm0.03}$  & --& --&--                                                          & $779/788$ & $10.7$ &$3.82$ & $6.88$ & --& --\\
81 & HY3302 & 61100.92 & -- & -- & -- & -- &        $0.79^{\pm0.01}$ & $2.53^{\pm0.03}$ & $40^{f}$       & $0.49^{\pm0.03}$  & --& --&--                                                          & $893/870$ & $10.8$ &$3.97$ & $6.83$ & --& --\\
82 & HY3603 & 61101.77 & -- & -- & -- & -- &        $0.75^{\pm0.02}$ & $2.42^{\pm0.06}$ & $40^{f}$       & $0.44^{\pm0.04}$  & --& --&--                                                          & $685/766$ & $9.27$ &$3.61$ & $5.66$ & --& --\\
83 & HY3701 & 61102.42 & -- & -- & -- & -- &        $0.77^{\pm0.03}$ & $2.58^{\pm0.08}$ & $40^{f}$       & $0.52^{\pm0.06}$  & --& --&--                                                          & $713/693$ & $10.2$ &$3.74$ & $6.46$ & --& --\\
84 & HY3702 & 61102.57 & -- & -- & -- & -- &        $0.79^{\pm0.01}$ & $2.49^{\pm0.05}$ & $40^{f}$       & $0.43^{\pm0.03}$  & --& --&--                                                          & $829/843$ & $10.4$ &$3.66$ & $6.74$ & --& --\\
85 & HY3703 & 61102.69 & -- & -- & -- & -- &        $0.80^{\pm0.02}$ & $2.52^{\pm0.07}$ & $40^{f}$       & $0.47^{\pm0.05}$  & --& --&--                                                          & $744/767$ & $10.1$ &$3.57$ & $6.53$ & --& --\\
86 & HY3401 & 61103.20 & -- & -- & -- & -- &        $0.80^{\pm0.02}$ & $2.47^{\pm0.05}$ & $40^{f}$       & $0.45^{\pm0.03}$  & --& --&--                                                          & $767/798$ & $10.1$ &$3.62$ & $6.38$ & --& --\\
87 & HY3801 & 61105.92 & -- & -- & -- & -- &        $0.79^{\pm0.01}$ & $2.46^{\pm0.05}$ & $40^{f}$       & $0.33^{\pm0.02}$  & --& --&--                                                          & $849/841$ & $9.24$ &$2.79$ & $6.45$ & --& --\\
88 & HY3802 & 61106.12 & -- & -- & -- & -- &        $0.79^{\pm0.01}$ & $2.44^{\pm0.05}$ & $40^{f}$       & $0.35^{\pm0.02}$  & --& --&--                                                          & $880/833$ & $9.30$ &$2.97$ & $6.33$ & --& --\\
89 & HY3901 & 61106.71 & -- & -- & -- & -- &        $0.78^{\pm0.01}$ & $2.51^{\pm0.05}$ & $40^{f}$       & $0.44^{\pm0.03}$  & --& --&--                                                          & $852/866$ & $9.03$ &$3.17$ & $5.86$ & --& --\\
90 & HY3902 & 61106.91 & -- & -- & -- & -- &        $0.77^{\pm0.01}$ & $2.48^{\pm0.04}$ & $40^{f}$       & $0.49^{\pm0.03}$  & --& --&--                                                          & $845/848$ & $8.97$ &$3.48$ & $5.49$ & --& --\\
 91 &  NZ1012 &  61117.50 &  -- &  -- &  -- &  -- &         $0.85^{\pm0.01}$ &  $1.96^{\pm0.05}$ &  $25^{f}$       &  $0.02^{\pm0.002}$  &  $6.42^{\pm0.19}$ &  $1.85^{\pm0.20}$ & --                              &  $198/201$ &  $4.78$ & $0.31$ &  $4.45$ &  $0.01$&  --\\
 92 &  NZ1014 &  61159.30 &  $0.45^{\pm0.01}$ &  $6.03$ &  $15.9$ &  C &  $0.84^{\pm0.03}$ &  $1.67^{\pm0.02}$ &  $22^{\pm6.15}$       &  $0.69^{\pm0.05}$  &  $6.15^{\pm0.15}$ &  $2.42^{\pm0.16}$ & --                                            &  $364/362$ &  $1.06$ &  $0.87$ &  $0.17$ &  $0.01$ &  --\\                                                                    
 \hline

\end{tabular}
\end{table*}

\end{document}